\documentclass[10pt,sigconf]{acmart}

\makeatletter
\def\@copyrightspace{\relax}
\makeatother

\usepackage{ulem}
\newcommand{\ra}[1]{\renewcommand{\arraystretch}{#1}}
\usepackage{color}
\usepackage{booktabs} % For formal tables
\usepackage{graphics}
\usepackage{graphicx} 
\usepackage[T1]{fontenc}
\usepackage[utf8]{inputenc}
\usepackage{subfigure}
\usepackage{balance}
\usepackage{amsmath}
\usepackage{colortbl}
\usepackage{tabularx}
\usepackage{xspace}
\usepackage{graphicx}    % For importing graphics
\usepackage{url} 
\usepackage{natbib}   
\usepackage{threeparttable}     
\usepackage{multirow}
\usepackage{caption}

\newcommand{\systemName}{SolarGest\xspace}

\setcopyright{none}

\begin{document}
\thanks{\textbf{*This is a pre-print version to appear at MobiCom 2019}}
\title{SolarGest: Ubiquitous and Battery-free Gesture Recognition using Solar Cells}

\author{Dong Ma}
\affiliation{\institution{UNSW \& Data61-CSIRO, Australia}}
\email{dong.ma1@unsw.edu.au}
\author{Guohao Lan}
%\authornote{Currently as a postdoc at Duke University.}
\affiliation{	\institution{University of New South Wales}}
\email{guohao.lan@unsw.edu.au}
\author{Mahbub Hassan}\affiliation{
	\institution{University of New South Wales}}
\email{mahbub.hassan@unsw.edu.au}
\author{Wen Hu}
\affiliation{	\institution{University of New South Wales}}
\email{wen.hu@unsw.edu.au}
\author{Mushfika B. Upama}
\affiliation{	\institution{University of New South Wales}}
\email{m.upama@unsw.edu.au}
\author{Ashraf Uddin}
\affiliation{	\institution{University of New South Wales}}
\email{a.uddin@unsw.edu.au}
\author{Moustafa Youssef}
\affiliation{	\institution{Alexandria University}}
\email{moustafa@alexu.edu.eg}

\renewcommand{\shortauthors}{D. Ma et al.}

\begin{abstract}
We design a system, SolarGest, which can recognize hand gestures near a solar-powered device by analyzing the patterns of the photocurrent. SolarGest is based on the observation that each gesture interferes with incident light rays on the solar panel in a unique way, leaving its distinguishable signature in harvested photocurrent. Using solar energy harvesting laws, we develop a model to optimize design and usage of SolarGest. To further improve the robustness of SolarGest under non-deterministic operating conditions, we combine dynamic time warping with Z-score transformation in a signal processing pipeline to pre-process each gesture waveform before it is analyzed for classification. We evaluate SolarGest with both conventional opaque solar cells as well as emerging see-through transparent cells. Our experiments with 6,960 gesture samples for 6 different gestures reveal that even with transparent cells, SolarGest can detect 96\% of the gestures while consuming 44\% less power compared to light sensor based systems.
\end{abstract}

%\keywords{Light-based Sensing, Visible Light Sensing, Gesture Recognition,}

\settopmatter{printacmref=false} % Removes citation information below abstract
\renewcommand\footnotetextcopyrightpermission[1]{} % removes footnote with conference information in first column
\pagestyle{empty}

\pagestyle{plain}

\thispagestyle{plain}
\cfoot{\thepage}
\pagestyle{plain}
\cfoot{\thepage}
\maketitle

\section{Introduction}
\subsection{Motivation}

\begin{figure}[]
	\centering
	\includegraphics[scale=0.45]{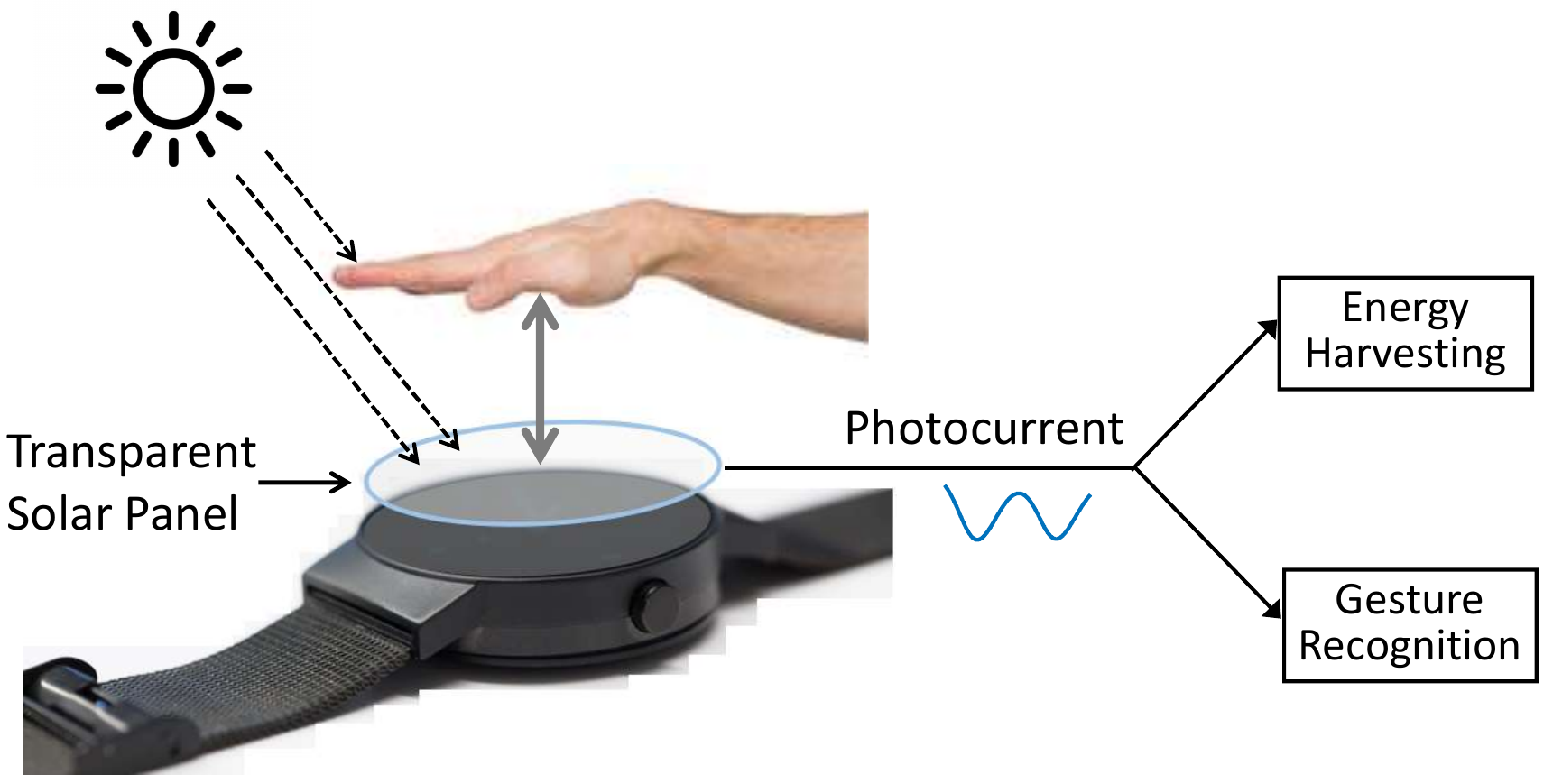}	
%	\vspace{-0.1in}	
	\caption{Illustration of a transparent solar powered smartwatch with solar-based gesture recognition.}
	\label{fig:solarWatch}
	\vspace{-0.15in}
\end{figure}

As all types of devices around us become smart and capable of taking input from us, we need to explore more natural ways to interact with them. There is a growing trend to integrate gesture recognition  to consumer electronics \cite{chaudhary2013intelligent,ren2011robust}, because it is one of the most natural ways for human to communicate with anyone or anything. Given the diversity of devices, many of which would be powered by small batteries, we need gesture systems that work with any device and consume zero energy in addition to the normal device operation. By using solar panels, we can achieve these two goals simultaneously, i.e., any device fitted with solar panels for energy harvesting can also recognize gestures. Since solar energy harvesting responds to any form of light, SolarGest can find applications both indoor and outdoor. For example, users can purchase from solar-powered vending machines, configure solar-powered garden lights, or operate solar-powered calculators by simply using gestures.

There is a new development in solar technology, \textit{transparent} solar cells \cite{wang2008transparent,traverse2017emergence}, which makes solar panels more attractive for mobile devices. Made from novel organic materials, transparent cells absorb and harvest energy from infrared and ultraviolet lights,
but let the visible lights pass through so we can see through the solar panel like a clear glass. With the discovery of transparent cells, solar panels now can be fitted to the entire device body, including on top of the screen, to harvest more energy. Figure \ref{fig:solarWatch} illustrates how a transparent solar cell fitted on the screen of a smart watch can be used for the dual purpose of energy harvesting as well as gesture recognition.

\subsection{Limitations of Existing Work}

There is a growing trend in exploring gesture systems for consumer devices using a variety of sensors and modalities, such as WiFi (electromagnetic) \cite{abdelnasser2015wigest,sbirlea2013automatic,pu2013whole}, camera (image) \cite{izadi2011kinectfusion,howe2000bayesian}, microphone (acoustic) \cite{gupta2012soundwave,pittman2016multiwave}, accelerometer (motion) \cite{ruiz2011user,xu2012taplogger}, and light sensor (ambient light) \cite{venkatnarayan2018gesture,kaholokula2016reusing,li2017reconstructing,li2015human,li2016practical}. Some of them, such as WiFi and accelerometer, are more ubiquitous than others and most of them can achieve high gesture accuracies up to 98\%. However, none of them harvest, but consume energy. Work on solar-based gesture is rare with the exception of a recent work by Varshney et al. \cite{varshney2017battery} that has experimentally demonstrated the feasibility of gesture recognition with a specific silicon-based \textit{opaque} solar cell. The findings from the preliminary work in \cite{varshney2017battery} open the door for ubiquitous solar-based gesture recognition for future Internet of Things (IoT), but it is limited in following ways. 

First, \cite{varshney2017battery} differentiates only three gestures based on the number of times the user repeats a basic hand movement, which is basically recognition of one gesture but with different counts. Although this method can be easily implemented using a simple threshold-based algorithm with a counter, it requires the user to remember the hand movement counts to ensure correct gesture is communicated. %Work in \cite{varshney2017battery} did not provide a framework for solar panels to recognize \textit{arbitrary} user-friendly gestures that do not rely on counting. 

The second limitation is lack of a theoretical model to simulate solar gesture recognition under different parameters. For example, we do not know how to analyze gesture recognition performance of solar cells as a function of lighting condition, efficiency and form factor of the solar cell, user hand size, and proximity of hand to the solar panel. Without a simulation model, design optimization of user-friendly solar gesture systems can be exhausting as one has to experimentally estimate the performance of the system for a large combination of parameter values. For example, how transparency of a solar cell may affect gesture recognition performance cannot be studied without first acquiring a series of transparent solar cells of specific properties, which can be very expensive,  limiting the possibility of future research in the area to explore new algorithms.

\subsection{Proposed Methodology}

We propose SolarGest, which detects user-friendly gestures of arbitrary design. It is based on the observation that any hand gesture interferes with incident light rays on the solar panel in a unique way, leaving its distinguishable signature in harvested photocurrent's time series data. By delegating the learning and detection responsibilities to machine learning, we can focus on designing user-friendly gestures beyond the simple counting-based gestures. 

We observe that any influence from hand gesture on the photocurrent is governed by solar energy harvesting laws, which can provide a quantitative estimation of the generated photocurrent given the intensities and incident angles of ambient lights, and the form factor as well as the energy harvesting density of the solar panel. Hand gestures change the volumes and angles of incident lights in a specific pattern, which can be explained using basic geometry. Combining solar energy harvesting law with geometry, we propose a model to simulate photocurrent waveforms produced by arbitrary hand gestures. A key utility of the model is that the future designers of SolarGest system can estimate gesture recognition performance of different types of solar cells for arbitrary gestures under different lighting environments before committing to costly experiments. Actual experiments can be done sparingly only for fine tuning the system.

%\subsection{Challenges and Solutions}

%In realizing SolarGest, we faced several challenges. Our model revealed that both duration and amplitude of the waveform can vary significantly from sample to sample for the same gesture due to variations in a number of hardware, environmental and user parameters such as solar cell form factor, intensity of ambient light, hand size, hand angle, speed of hand motion, and proximity of hand to solar panel. These variations make it challenging to train classifiers for accurate detection of specific gestures. We designed a combination of dynamic time warping (DTW) and z-score transformation to pre-process all gesture wave forms to reduce these variations before they are used by the classifier. As a result of this, we were able to achieve high gesture recognition accuracy even with basic machine learning.
%
%Validating the model was another challenge, especially for the emerging transparent solar cells, which are currently not available off-the-shelf. Using organic material, we developed two transparent solar cells of different levels of transparency and energy harvesting density in our own lab. 

%[needs enhancements: was there any challenge in developing the model? any other challenge we can highlight? ]    

\subsection{Contributions}
In realizing SolarGest, we faced several challenges. Our model revealed that both duration and amplitude of the waveform can vary significantly from sample to sample for the same gesture due to variations in a number of hardware, environmental and user parameters such as solar cell form factor, intensity of ambient light, hand size, hand angle, speed of hand motion, and proximity of hand to the solar panel. These variations make it challenging to train classifiers for accurate detection of specific gestures. We designed a combination of dynamic time warping (DTW) and Z-score transformation to pre-process all gesture waveforms to reduce these variations before they are used by the classifier. As a result of this, we were able to achieve high gesture recognition accuracy even with basic machine learning. Validating the model was another challenge, especially for the emerging transparent solar cells, which are currently not available off-the-shelf.
\begin{figure*}[]
	\centering
	\subfigure[Incident light.]{	
				\centering
				\includegraphics[scale=0.46]{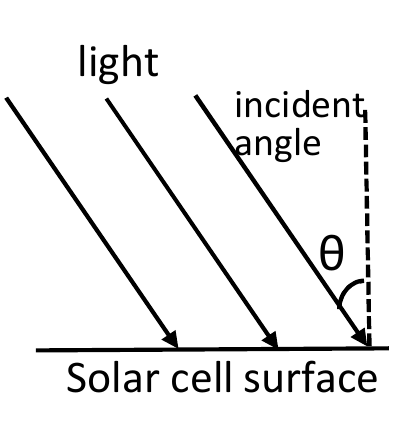}		
			\label{fig:angle}}
	\subfigure[3D geometric model.]{	
			    \centering
				\includegraphics[scale=0.39]{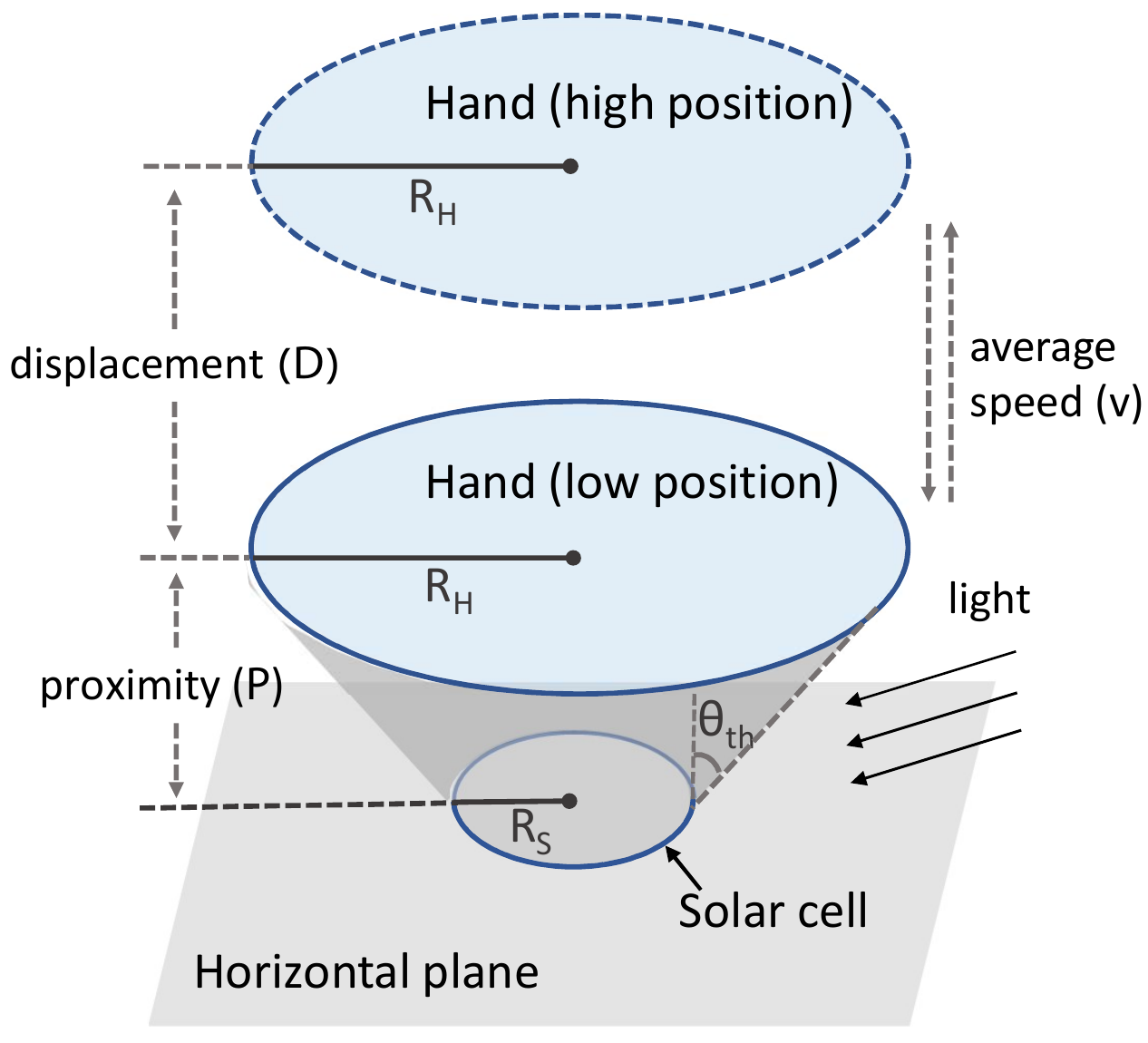}		
			\label{fig:3d}} %45
	\subfigure[2D analysis - vertical movement.]{	
			    \centering
				\includegraphics[scale=0.31]{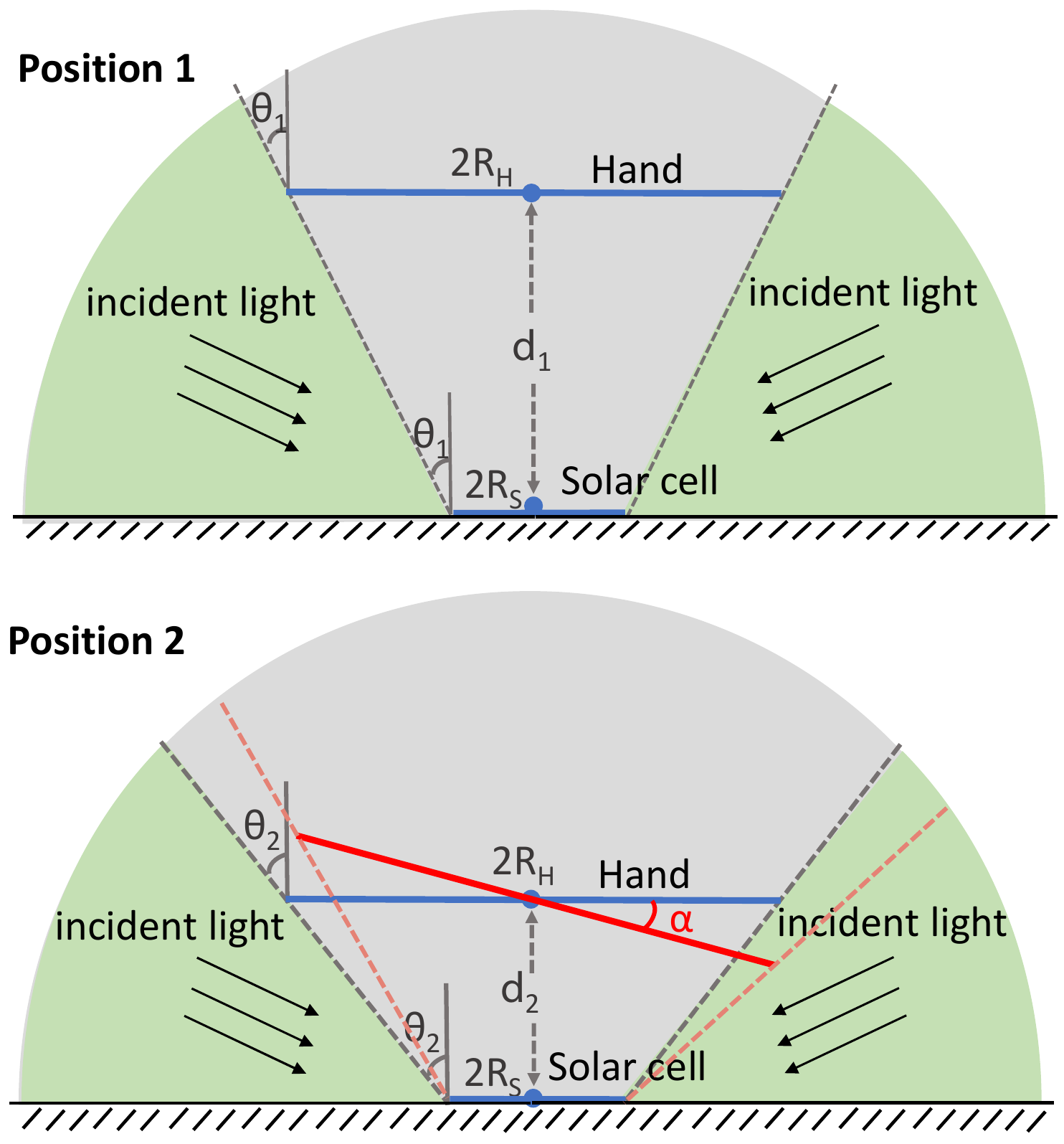}		
			\label{fig:vertical}}
	\subfigure[2D analysis - horizontal movement.]{	
			\centering
			\includegraphics[scale=0.31]{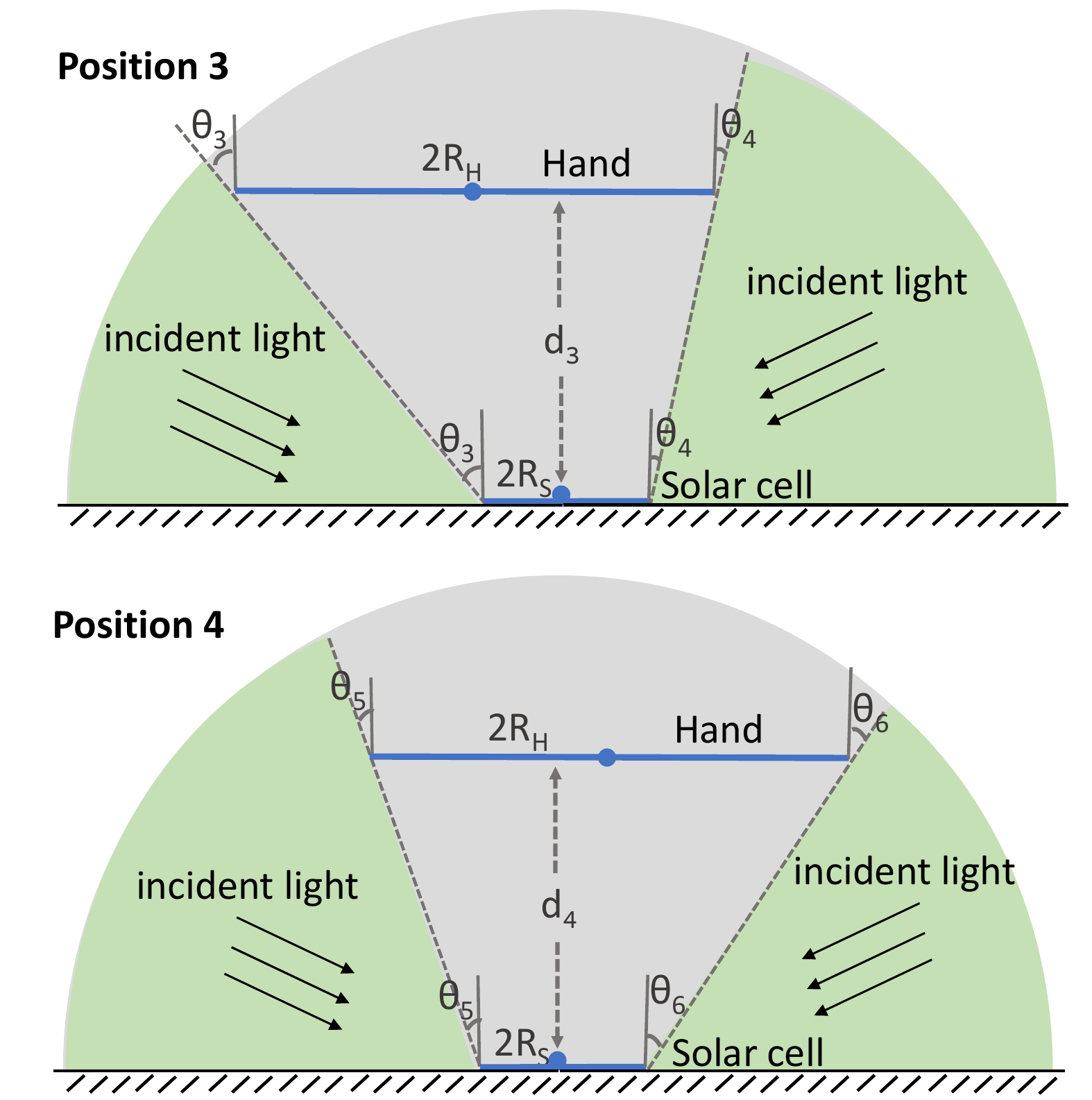}\label{fig:vertical1}} %35
%	\subfigure[Case 1]{	
%		    \centering
%			\includegraphics[scale=0.35]{fig/vertical1.pdf}		
%		\label{fig:vertical1}}
%	\subfigure[Case 2]{	
%		    \centering
%			\includegraphics[scale=0.35]{fig/vertical2.pdf}	
%		\label{fig:vertical2}}
%	\vspace{-0.1in}
	\caption{(a) Illustration of incident angle. (b) 3D geometric model of \systemName. (c) 2D geometric analysis of vertical movement. (d) 2D geometric analysis of horizontal movement.}
	\label{fig:v}
\end{figure*}

%\vspace{-0.1in}

%\sout{Validating the model was another challenge, especially for the emerging transparent solar cells, which are currently not available off-the-shelf. Using organic material, we developed two transparent solar cells of different levels of transparency and energy harvesting density in our photovoltaic lab. Key contributions of this paper can be summarized as follows:}

Key contributions of this paper can be summarized as follows:
\setlength{\leftmargini}{1.2em}
%\vspace{-0.1in}
\begin{itemize}
	\item  Using solar energy harvesting laws, we develop a model to simulate photocurrent waveforms produced by arbitrary hand gestures in both \textit{vertical} and \textit{horizontal} planes relative to the solar panel. The model allows us to analyze gesture recognition performance of solar cells as a function of important parameters such as lighting condition, efficiency and form factor of the solar cell, user hand size, and proximity of hand to the solar panel. Using practical examples, we illustrate how the model can be used to optimize design and usage of SolarGest (Section \ref{sec:simulation}). 
		
	\item We propose a general machine learning framework to detect any type of gestures. By combining discrete wavelet transform (DWT), dynamic time warping (DTW), and Z-score transformation, we design an end-to-end signal processing pipeline to protect SolarGest performance against variations in operating conditions (Section \ref{sec:solarGR}). 
	
	\item Using organic material, we developed two transparent solar cells of different levels of transparency and energy harvesting density in our photovoltaic lab. We conduct real experiments with both silicon-based opaque solar panels as well as see-through organic solar cells. With 6,960 gesture samples collected for six user-friendly gestures under different light conditions, we validate our model and demonstrate that even for transparent cells, SolarGest can detect gestures with an accuracy of 96\%, which is comparable to that achieved with light sensors (Section \ref{sec:evaluation}). 
	
	\item Finally, we experimentally demonstrate that SolarGest consumes 44\% less power compared to systems that detect gestures using light sensors (Section \ref{s:power}). 
\end{itemize}

%The rest of the paper is organized as follows. Section~\ref{sec:simulation} presents the simulation model for solar-based gesture recognition, followed by the \systemName system design in Section~\ref{sec:solarGR}. Experimental evaluation of gesture recognition and power consumption are presented in Sections~\ref{sec:evaluation} and \ref{s:power}, respectively. We review related work in Section~\ref{sec:related} before concluding in Section~\ref{sec:conclusion}. 

%\begin{figure}[t]
%	\centering
%		\includegraphics[scale=0.6]{fig/angle.pdf}		
%	\caption{Illustration of direct light and oblique light.}
%	\label{fig:angle}
%\end{figure}
%

\section{Solar Gesture Simulator}
\label{sec:simulation}
Using fundamentals of solar energy harvesting and simple geometrical arguments, we derive a model to simulate photocurrent waveforms produced by hand gestures containing arbitrary hand movements in both \textit{vertical} and \textit{horizontal} planes relative to the solar panel. The model allows us to study the impact of different system parameters such as lighting condition, efficiency and form factor of the solar cell, etc., on the photocurrent waveform. Using numerical experiments, we illustrate the utility of the model in terms of predicting gesture recognition accuracy and optimizing the design and usage of solar-based gesture recognition systems. 

\subsection{Modeling Solar Gestures}
\label{s:model}
%\subsection{Fundamentals of Solar Energy Harvesting}
Due to photovoltaic effect~\cite{parida2011review}, solar cells convert incident light energy into electrical current (photocurrent). The amount of photocurrent generated is a function of the \textit{form factor} of the solar cell and its \textit{current density}, i.e., the amount of photocurrent generated per unit area (e.g., $mA/cm^2$), which is a measure of solar energy harvesting efficiency and depends on the light intensity of the operating environment. To fairly compare the efficiency of different solar cells, current density is typically reported under a standard lighting condition, named Global Standard Spectrum (AM1.5g)~\cite{smestad2008reporting,riordan1990air}. Then, the standard current density $J^*_{SC}$ is obtained as~\cite{wright2012organic}:

\begin{equation}
J^*_{SC}=\frac{q}{h c_{0} } \int_0^ \infty {a( \lambda )I(\lambda )\lambda d\lambda} 
\label{eq:1}
\end{equation}
where $q$ is the elementary charge, $c_{0}$ is the speed of light in free space, and $h$ is the Planck's constant. Symbol $\lambda$ refers to the wavelength of incident light. $a(\lambda)$ and $I(\lambda)$ represent the solar cell absorption efficiency and light intensity at wavelength $\lambda$, respectively. Due to the linear relationship between current density and light intensity~\cite{cai2011effect}, one can calculate current density $J_{SC}$ ($mA/cm^2$) at any light intensity $I$ by, 
\begin{equation}
J_{SC}=\frac{I}{I^*} J^*_{SC} 
\label{eq:5}
\end{equation}
where $I^*(=100mW/cm^2)$ is the light radiance power under Global Standard Spectrum (AM1.5g). Then, using Lambert's cos($\theta$)-law~\cite{weik2000lambert}, generated photocurrent $J$, is obtained as 

%In practice, solar cells absorb energy from both direct light and oblique light, as shown in Figure~\ref{fig:angle}. The angle between light beams and surface normal is defined as the incident angle, denoted by $\theta$. As a result, suggested by the well-known Lambert's cos($\theta$)-law~\cite{weik2000lambert}, the generated current $J$ is calculated as 
\begin{equation}
 J=  \int_0^{\pi/2} {  S \cdot J_{SC} \cdot cos(\theta) d\theta}
 \label{eq:2}
 \end{equation}
where $S$ is the form factor of the solar cell and $\theta$ is the \textit{incident angle}, i.e., the angle between light beam and surface normal (see Figure~\ref{fig:angle}). Light from different sources, like sun, fluorescent lamp and LED, can have a different spectral irradiance profile, resulting in different amount of photocurrent even under the same light intensity. Although we derive the model based on AM1.5g, which is specifically for sunlight, it is applicable to other irradiance spectrum as gestures are differentiated due to their unique patterns, rather than the absolute values. This will be further validated in Figure~\ref{fig:simulGestureCompare}, which confirms that gesture patterns collected under fluorescent light are consistent with the modeled counterparts.

\begin{figure}[t]
	\centering
	\subfigure[]{
		\includegraphics[scale=0.285]{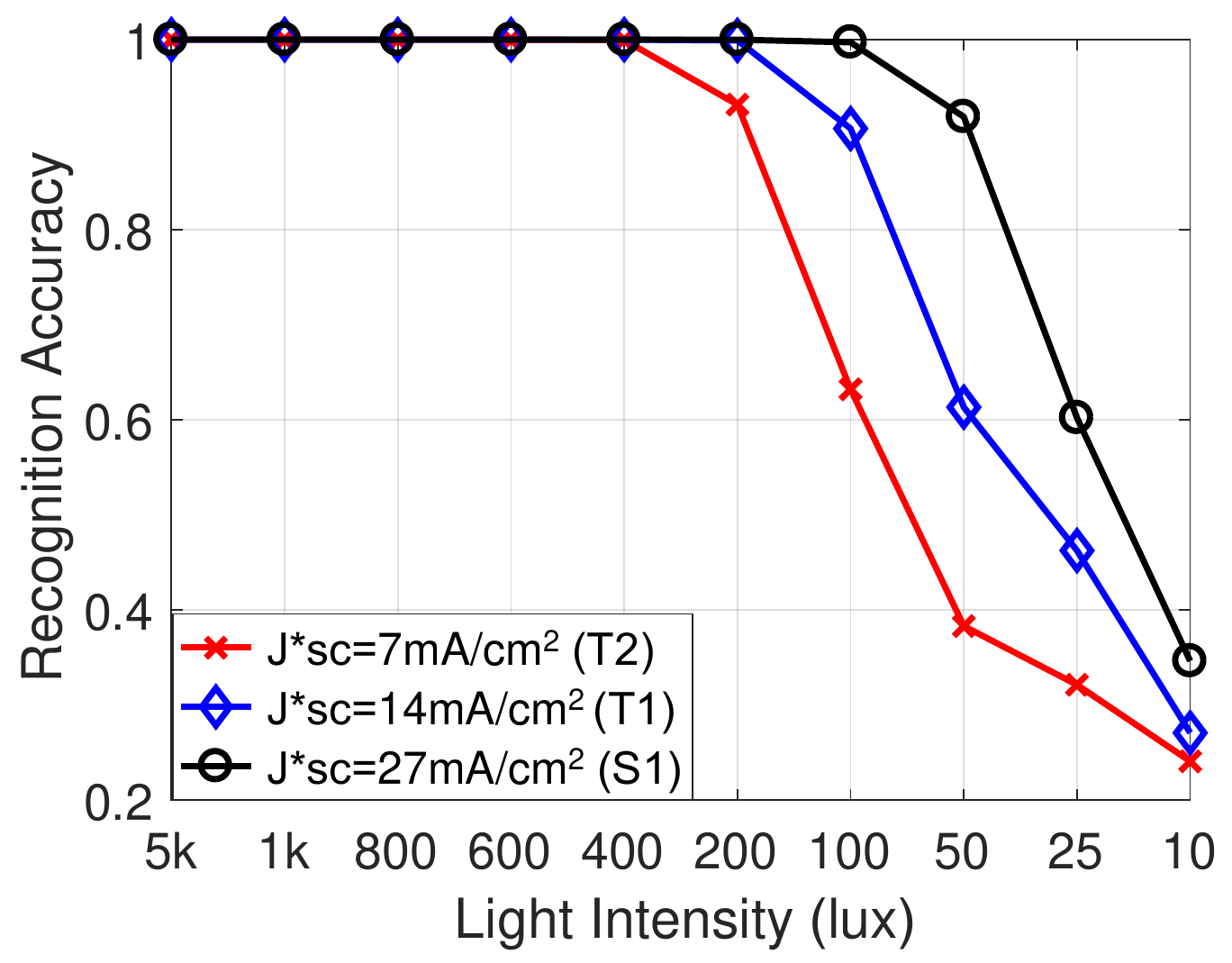}		
		\label{fig:simulExample2}}
	\subfigure[]{
		\includegraphics[scale=0.285]{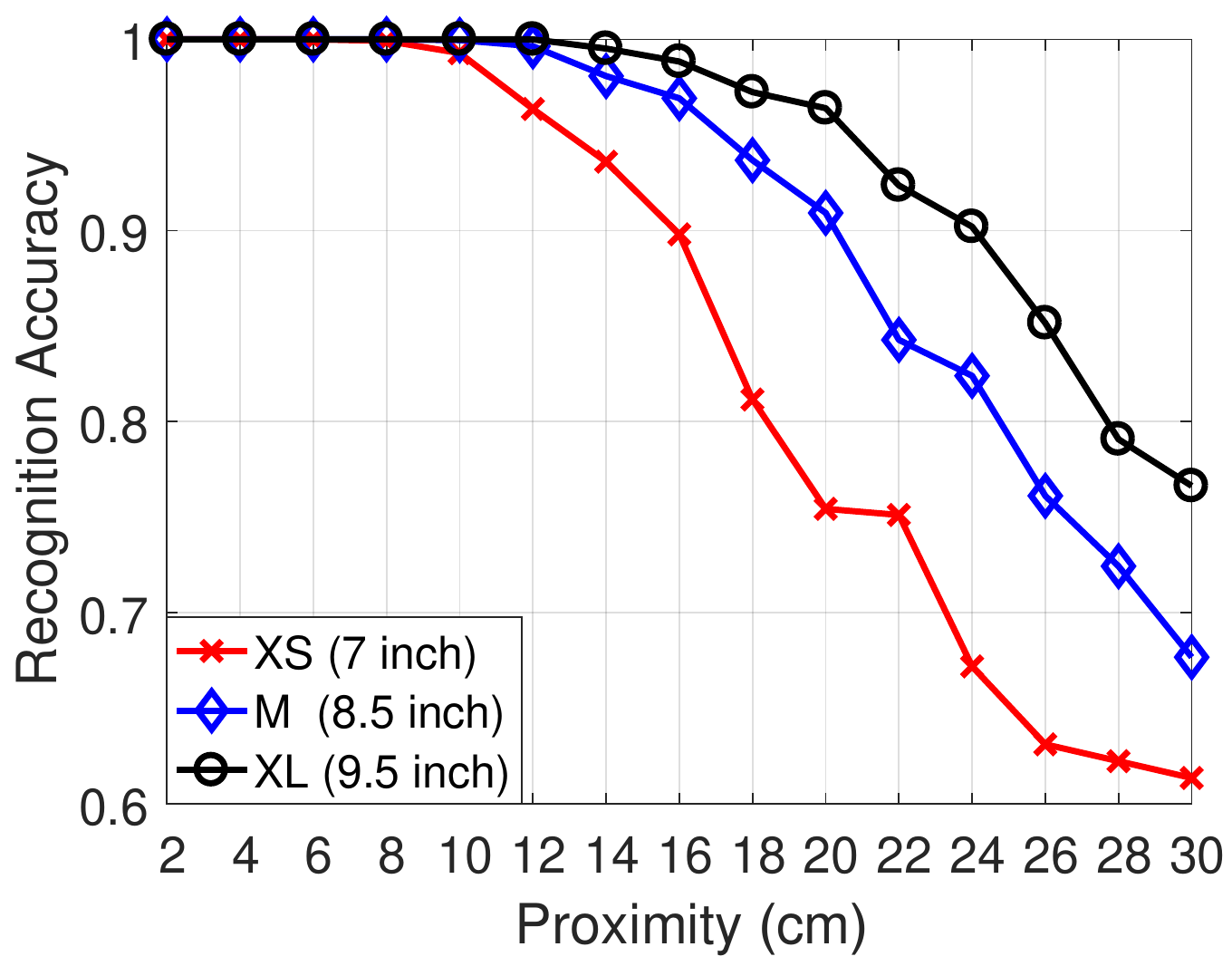}		
		\label{fig:simulExample1}}
		\vspace{-0.15in}
	\caption{(a) Simulated recognition accuracy versus light intensity for different energy harvesting efficiency. (b) Simulated recognition accuracy versus proximity for different hand size.}
	\vspace{-0.1in}
\end{figure}

%\subsection{Modeling Solar Energy Harvesting under Hand Gesture}
%\label{s:modeling}
%The underlying principle of \systemName is that hand gestures leave unique signature on solar cell current. Next, we derive the current profile under different gestures. 
To model solar photocurrent under hand gestures, we present a 3D geometric model, as shown in Figure~\ref{fig:3d}, in which human hand and solar cell are modeled as round surfaces with radius $R_H$ and $R_S$, respectively. As many IoT devices have small form factors~\cite{blaauw2014iot}, in this paper, we consider the case where solar cell is smaller than hand size (e.g., Lunar Watch~\cite{lunar}), i.e., $R_S<R_H$ (note that the model can be easily extended $R_S>R_H$~\footnote{In this case, the inner part of the solar cell, a circle with radius $R_H$, will be affected by hand movement, but the residual area will generate steady photocurrent during a gesture. Thus, total photocurrent would be obtained as the sum of current from the two parts.}).  The solar cell is assumed to be placed on a horizontal surface and a hand performs different gestures in a parallel plane above it. During a gesture, we define the minimum distance between the solar cell and hand as \textit{proximity}, denoted by $P$, and define the vertical movement space as \textit{displacement}, denoted by $D$. Since $R_S<R_H$, only the light rays with incident angles larger than a certain threshold ($\theta_{\textrm{th}}$) can hit the solar cell. 
%In detail, \textit{if the incident angle $\theta \in (0^{\circ},\theta_{th})$, the light beam is blocked; otherwise, if $\theta \in {[}\theta_{th}, 90^{\circ})$, the light beam can be absorbed by the solar cell.} Using the longitudinal section analysis, we next derive solar current under hand gestures.

%\begin{figure}[t]
%	\centering			
%	\includegraphics[scale=0.5]{fig/plotNoiseDistribution.pdf}
%	\caption{Noise distribution of real solar cell signals. The red curve plots the standard Gaussian distribution.}
%	\label{fig:plotNoiseDistribution}
%\end{figure}

%\subsubsection{Specific Hand Position }
Figure~\ref{fig:vertical} and (d) show the longitudinal section of the 3D model, in which solar cell and hand are represented by two line segments with lengths $2R_S$ and $2R_H$, respectively. The \textit{green area} indicates the angular space in which light can be absorbed by the solar cell, while the light in \textit{gray area} is blocked. In fact, a gesture is comprised of a time series of hand positions. Given the initial hand position, moving direction and speed of hand movement, one can calculate hand positions at any successive points in time. Taking \textit{Up} gesture as an example, if the initial distance (at time zero) between hand and solar cell is $d$ and the hand moves in a constant speed $v$, at time $t$, the distance between hand and solar cell becomes $d+vt$. Thus, the corresponding threshold angles $\theta_\textrm{th1}(t)$ and $\theta_\textrm{th2}(t)$ for the two absorption angular spaces are 
\begin{equation}
\theta_\textrm{th1}(t)=\theta_\textrm{th2}(t)=\arctan(\frac{R_H-R_S}{d+vt})
\label{eq:3}
\end{equation}
Since only light beams from the two green areas can be absorbed, the photocurrent $J{(t)}$ can be calculated as

\begin{equation}
J(t)=  \int_{\theta_\textrm{th1}(t)}^{\pi/2} {  S \cdot J_{SC} \cdot cos(\theta) d\theta} +  \int_{\theta_\textrm{th2}(t)}^{\pi/2} {  S \cdot J_{SC} \cdot cos(\theta) d\theta}
\label{eq:4}
\end{equation}
From Eq.\ref{eq:4}, the complete gesture waveform can be obtained by generating photocurrent values at successive points in time, i.e., $(J(t_1),J(t_2),...,J(t_n))$, where $J(t_1)$ and $J(t_n)$ represent the start and end of the gesture, respectively. Finally, presence of noise can be easily modeled by adding a noise term to each sample as $(J(t_1)+\epsilon,J(t_2)+\epsilon,...,J(t_n)+\epsilon)$.

\subsection{Estimating Recognition Performance}
\label{ss:estimating}
%\subsection{Model Utility}
%\subsubsection{Validity of the Modeled Gesture}
%To evaluate the validity of our model, we collected some gesture data using two real transparent solar cells T1 and T2, with current density of $14mA/cm^2$ and $7mA/cm^2$. (presented in Section~\ref{sec:data acquisition}). As shown in Figure~\ref{fig:simulGestureCompare}, the graphs in the first row illustrate profiles of five gestures simulated by our model, in which the two curves represent the same gesture generated by different solar cells. Graphs in the second row are the real collected signals. It can be observed that, gesture profiles created by our model are very similar to that generated by real solar cells. Thus, our model can be an effective tool to study the performance of \systemName under a variety of scenarios that are either impossible or exhaustive in practice. Next, we present two examples of using our model to facilitate the design and usage of \systemName.

\begin{figure*}[t]
	\centering
		\includegraphics[scale=0.6]{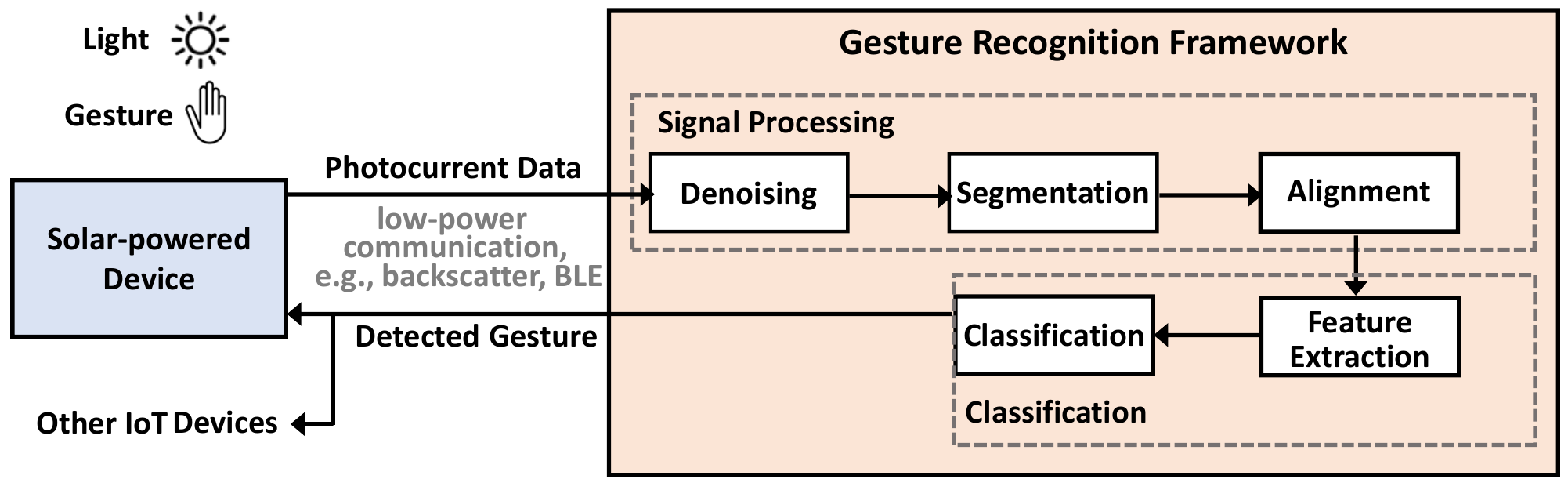}		\vspace{-0.1in}
	\caption{SolarGest System Architecture.}
	\label{fig:pipeline}
	\vspace{-0.1in}
\end{figure*}

Using the equations derived in Section~\ref{s:model}, our model is able to simulate gestures under various conditions, such as varying light intensity, proximity, as well as energy harvesting efficiency ($J^*_{SC}$). For the same gesture, we can also generate many synthetic samples by simulating human imperfection, such slight variations in speed, proximity, displacement, etc., or hand size variations of a family of users. These simulated gesture samples, on the one hand, can be used by future researchers to explore and compare different gesture recognition algorithms. On the other hand, solar-powered IoT designers can use these synthetic gesture samples to analyze and optimize various design tradeoffs. Next, we simulate 5 different gestures: \textit{Up, Down, UpDown, DownUp, LeftRight} (see Figure~\ref{fig:simulGestureCompare}), and utilize the  gesture recognition framework proposed in Section~\ref{sec:solarGR} to illustrate such trade-off analysis using numerical experiments.

For all numerical experiments, we set the following default parameter values:  $R_H=6cm$, $R_S=2cm$, $D=12cm$ and $P=3cm$. Solar cell current density $J^*_{SC}$ under the Global Standard Spectrum (AM1.5g) is set to $7mA/cm^2$ and the average speed of a gesture is set to $20cm/s$. The light intensity is set to 5000 lux. By randomly varying the speed of hand motion and proximity, we generate 100 random samples for each of the five gestures.

Figure~\ref{fig:simulExample2} plots simulated recognition accuracy as a function of light intensity, for three different solar energy harvesting efficiencies or different transparency levels. Here T1 and T2 denote transparent solar cells and S1 denotes opaque cell (see details of our solar cell prototyping in Section \ref{ss:prototype}). We can observe that, solar cells with different transparency levels require different lighting conditions to achieve a target gesture recognition accuracy. For example, to achieve 90\% accuracy in relatively dim lighting environment (50 lux), we must use a less transparent cell ($27mA/cm^2$), but a highly transparent cell ($7mA/cm^2$) can be used for improved visibility if the typical operating environment is well lit (200 lux).

Another interesting observation from Figure~\ref{fig:simulExample2} is that, according to our model, transparent solar cells (both T1 and T2) are predicted to recognize hand gestures with very high accuracy (close to 100\%) just like the opaque cells (S1) as long as the light intensity is higher than 400 lux. Given that many indoor and even very cloudy outdoor environments enjoy light intensities above 400 lux, this numerical experiment suggests that use of transparent solar cells will not have any negative effect on gesture recognition for typical use scenarios of future transparent solar-powered IoTs. Indeed, our practical experiments in Section \ref{sec:evaluation} with both opaque and transparent solar cells will validate this finding.

%\subsubsection{Utility to User}
For different hand sizes, Figure~\ref{fig:simulExample1} plots simulated recognition accuracy as a function of gesture proximity. The results indicate that, to achieve a certain recognition accuracy, users with smaller hand size should perform gestures closer to the solar panel, compared to those with larger hand size. Such numerical experiments can be used by solar-powered IoT manufacturers to release gesture guidelines for different hand sizes, which would help all users of the product to enjoy high gesture recognition accuracies.

 \section{Recognition Framework}
 \label{sec:solarGR}

Figure~\ref{fig:pipeline} illustrates the system architecture and workflow of \systemName. During a hand gesture near it, a solar-powered device captures time-series of photocurrent and delivers the data to a gesture recognition system, which could be located on an edge device, such as smartphone, laptop, or home hub (note that such edge-based processing will ensure that latency is minimal), using low-power communications like backscatter or BLE.  The gesture recognition system detects the gesture and either sends that information back to the originating device if local control in the device is involved, or communicates with other IoT devices based on the desired action from the gesture. Recognition accuracy is the key performance measure for the gesture recognition system. We propose a machine learning based gesture recognition framework that trains a classifier with specific features extracted from the photocurrent time series of the gesture. Before extracting features, we pass the signal through a processing pipeline to deal with a number of issues that may cause high classification errors. Signal processing and classification details of our proposed gesture recognition framework are presented next.

 \subsection{Signal Processing}
 
The signal processing pipeline deals with three specific issues. First, it removes noise contained in photocurrent signal using discrete wavelet transform (DWT). Then, the boundaries of the gesture are detected using a segmentation algorithm. Finally, a signal alignment module applies a combination of dynamic time warping (DTW) and Z-score transformation on the segmented signal to address specific alignment issues that are caused by variations in operating conditions, such as hand motion speed, lighting conditions etc.

 \subsubsection{Denoising}
 \label{s:denoising}
\begin{figure}[t]
\centering
\subfigure[]{
	\includegraphics[scale = 0.66]{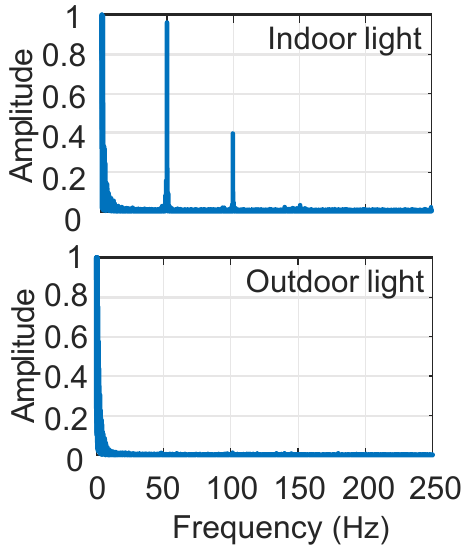}	
	\label{fig:FFT}}	
\subfigure[]{
	\includegraphics[scale = 0.342]{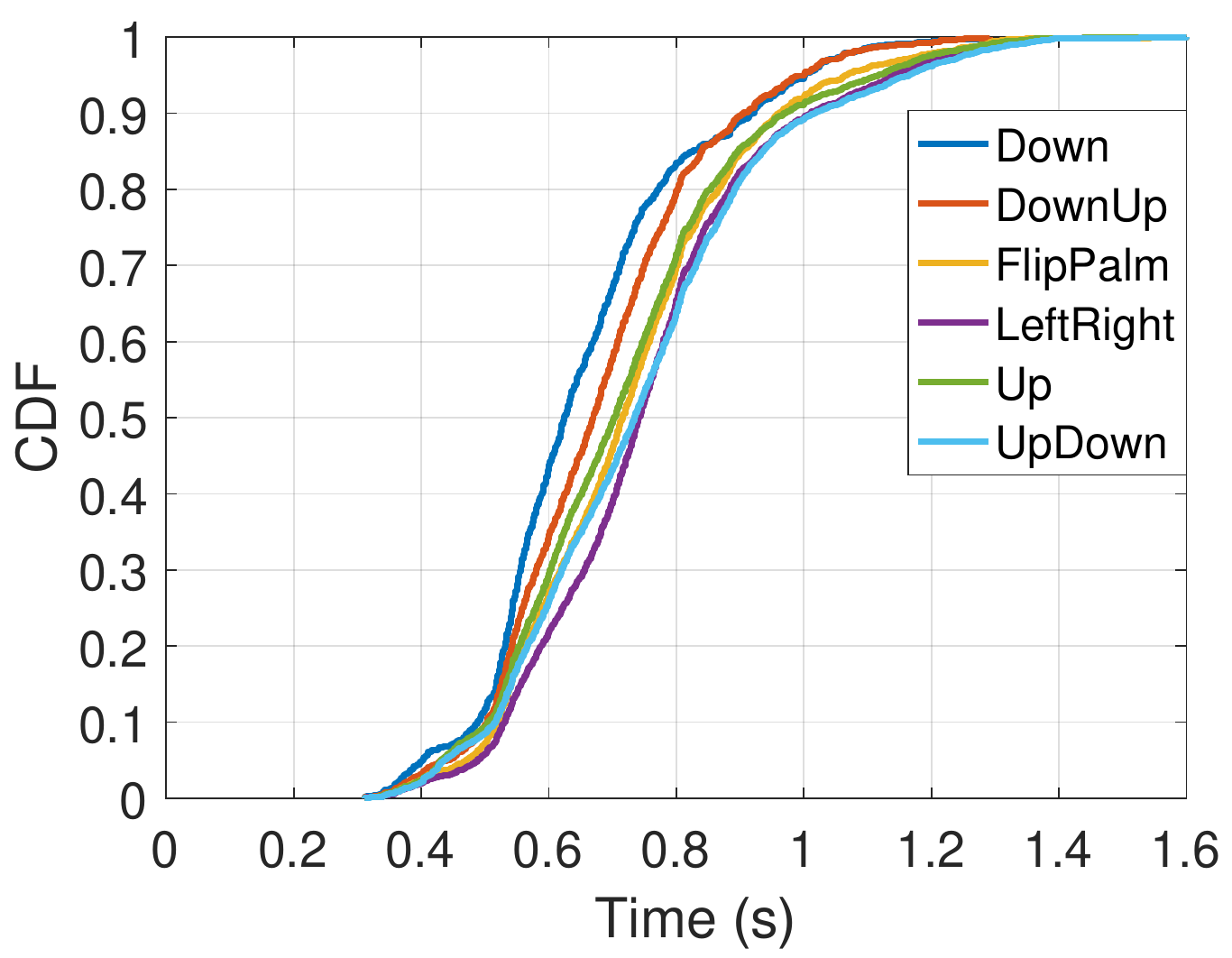}	
	\label{fig:plotLengthCDF}}
\vspace{-0.15in}
\caption{(a) FFT analysis of photocurrent signals collected under indoor fluorescent light (up) and outdoor natural light (bottom), (b) CDFs of gesture duration.}
\vspace{-0.19in}
%\label{fig:plotLengthCdfDF}
\end{figure}

Raw photocurrent signals are noisy as shown in the bottom row of Figure \ref{fig:simulGestureCompare}. The fast Fourier transform (FFT) graphs in Figure~\ref{fig:FFT} reveal that there is a 50Hz noise when the signal is collected indoor under a ceiling light powered by 50Hz AC current, but such noise is absent when measured outdoor under the sun. In addition, due to the minor imperfections in micro-controller of Arduino, Gaussian noise also exists in the photocurrent signal. Earlier works \cite{venkatnarayan2018gesture} have found that discrete wavelet transform (DWT) can effectively capture both temporal and frequency information, thereby filtering noise from both time and frequency domain~\cite{lang1996noise}. Unlike FFT that decomposes a signal in equal resolution over the whole frequency span, DWT is able to resolve a signal in various resolutions at different frequency range. More specifically, DWT hierarchically decomposes a signal and provides detail coefficients and approximation coefficients at each level. The key insight to denoise signal using DWT is to modify the detail coefficients based on thresholding strategies. Specifically, we divide the denoising procedure into three steps.

First, \systemName decomposes the photocurrent signal to level 5. The intuition to choose level 5 is based on the sampling rate. Since we sample data in 500Hz, the highest frequency contained in the signal is 250 Hz due to the Nyquist Theorem. As observed from Figure~\ref{fig:FFT}, the gesture frequency is actually less than around 5Hz. During DWT decomposition, the frequency span at each level is half of the level before it. Thus, at level 5, the frequency range is [0, $250/2^{5}$]Hz, i.e., [0, 7.8] Hz, which covers the gesture frequency. Second, a soft thresholding scheme is applied to the detail coefficients at level 5, which shrinks both positive and negative coefficients towards zero. The threshold is adaptively computed using the principle of Stein's Unbiased Risk Estimate (SURE) ~\cite{gradolewski2013use}. Finally, inverse DWT is applied to the altered detail coefficients and unmodified approximation coefficients to reconstruct the denoised signal. Due to space limitation, theory of DWT is not provided and readers can refer to~\cite{burrus1998introduction,lang1996noise} for more details.
 % More specifically, for given coefficient $d$ and threshold $\lambda$, if $d>\lambda$, $d$ becomes $d-\lambda$, and if $d<-\lambda$, $d$ becomes $d+\lambda$, otherwise, $d$ is set to 0.

%  \begin{figure}[t]
%  \centering
%  \subfigure[Two detected LeftRight gestures]{
%  	\includegraphics[scale=0.19]{fig/plotAlignment1.pdf}	
%  	\label{fig:plotAlignment1}}	
%  \subfigure[After Z-score Transformation]{
%  	\includegraphics[scale=0.19]{fig/plotAlignment2.pdf}	
%  	\label{fig:plotAlignment2}}
%  \subfigure[After Z-score Transformation and DTW]{
%  	\includegraphics[scale=0.19]{fig/plotAlignment3.pdf}	
%  	\label{fig:plotAlignment3}}
%  \caption{Illustration of signal alignment using Z-score transformation and DTW.}
%  %\label{fig:plotLengthCdfDF}
%  \vspace{-0.15in}
%  \end{figure} 

 \subsubsection{Gesture Segmentation}
 \label{s:gesture detection}

After denoising, the next step is to segment exact gestures from the time-series of signal. To help detect the start and stop of a gesture, like many other gestures recognition systems~\cite{virmani2017position,venkatnarayan2018gesture,kaholokula2016reusing,abdelnasser2015wigest}, \systemName requires users to take a short pause before and after a gesture. To detect the start and end of a gesture, previous works either use a preamble scheme~\cite{abdelnasser2015wigest} or a threshold-based method (i.e., a start is detected once the value is higher than a predefined threshold and an end is detected when the values fall below the threshold)~\cite{venkatnarayan2018gesture,kaholokula2016reusing}. However, the former requires users to perform an additional gesture every time, which is not user-friendly. The threshold-based method does not work if the amplitude before and after a gesture is different (e.g., \textit{Up} and \textit{Down} shown in Figure~\ref{fig:plotGestureIllustration}). Thus, we proposed a new segmentation algorithm, which can accurately detect the \textit{plateau periods} (i.e., pauses) before and after a gesture.

 \begin{figure}[t]
   	\centering
   	\includegraphics[scale = 0.6]{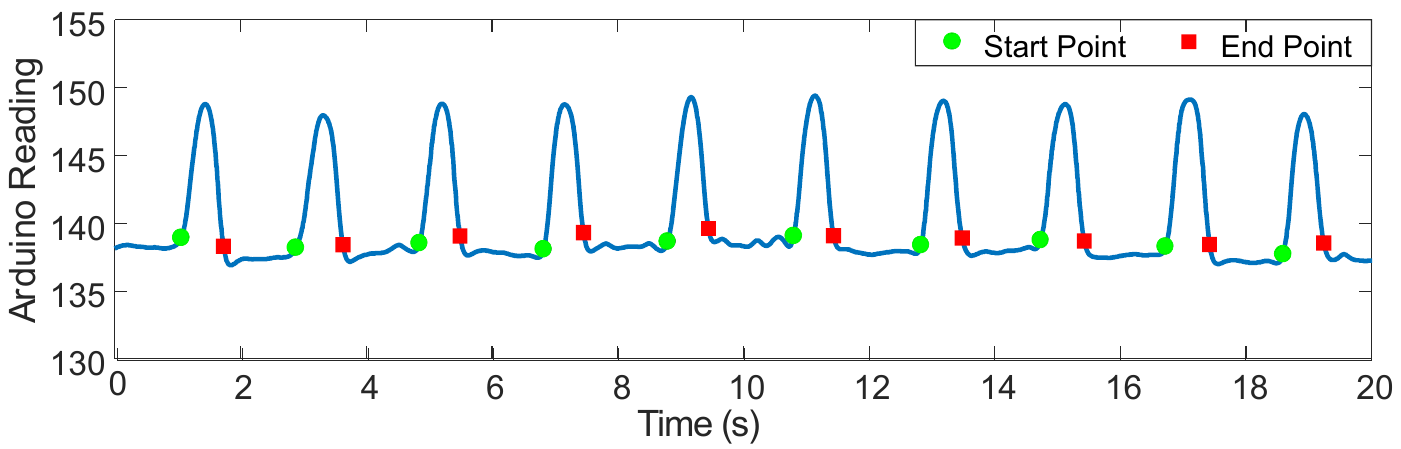}
   	\vspace{-0.25in}
   	\caption{Segmentation performance. The green dots represent the detected start points and the red squares represent the detected end points.}
   	\label{fig:SegmentationComparison1}
   	\vspace{-0.25in}
   \end{figure}
   
   \begin{figure*}[]
   	\centering
   		\includegraphics[scale=0.74]{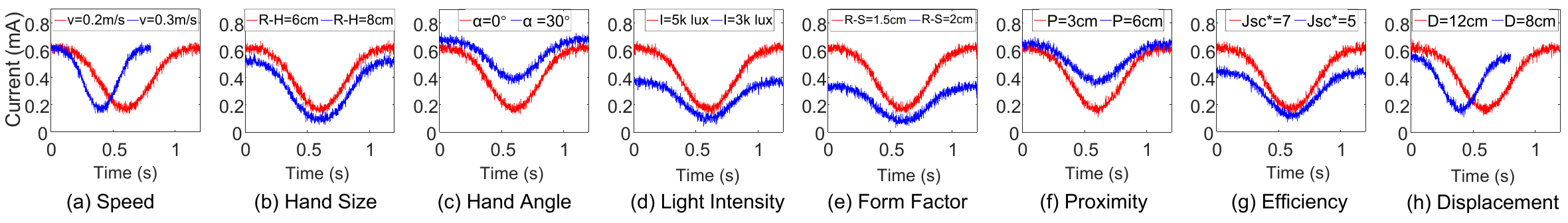}		
   		\vspace{-0.25in}
   	\caption{The impact of different parameters on gesture profile. In each graph, only a specific parameter varies and the rest are in default value.}
   	\label{fig:simulParameters}
   \end{figure*}
    \begin{figure}[t]
    \centering
    
    	\includegraphics[scale=0.35]{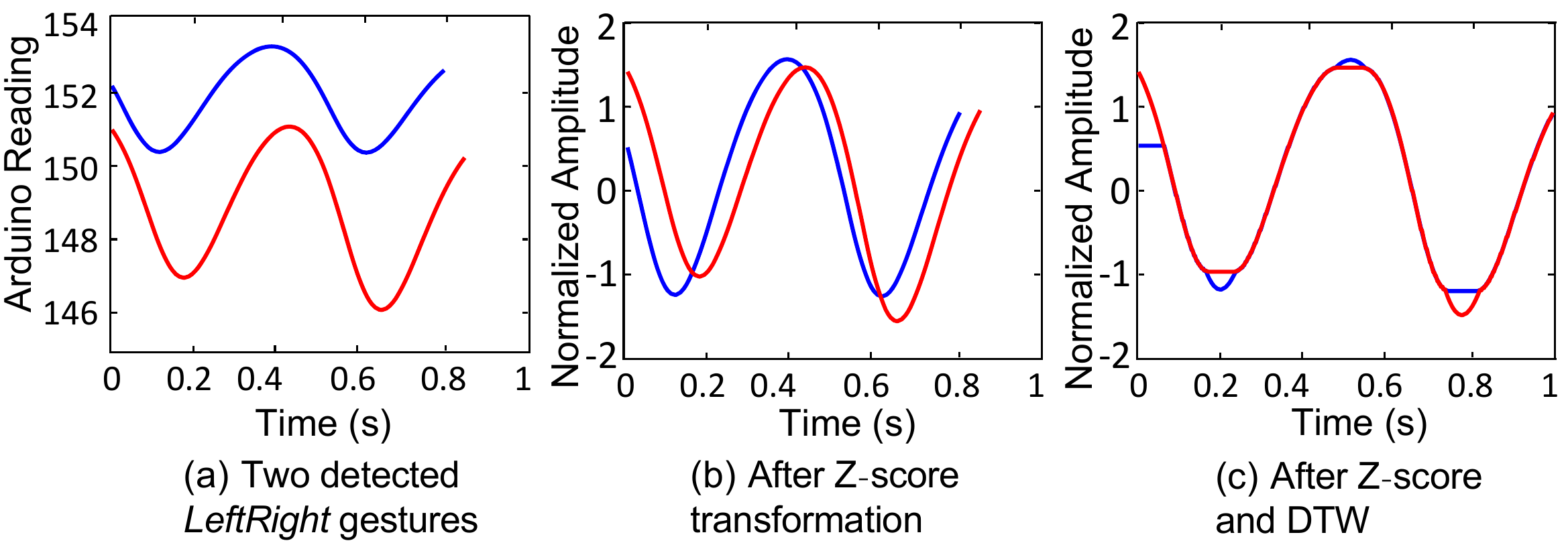}	
    		
   \vspace{-0.1in}
    \caption{Illustration of signal alignment using Z-score transformation and DTW.}
    \label{fig:plotAlignment}
    %\label{fig:plotLengthCdfDF}
    \vspace{-0.15in}
    \end{figure}

Specifically, we apply a sliding temporal window on the denoised signal. A gesture start is detected if the following two conditions hold: (1) the standard deviation of the samples in current window is lower than a pre-defined threshold \textit{stdThr}; (2) the difference between the last sample in current window and the mean of all the samples in the window is higher than a threshold \textit{meanThr}. The first condition ensures that the current window is in a plateau, while the second condition determines that a gesture starts right after a pause. Thus, the last sample in current window is regarded as a gesture start. The same principle is utilized to detect the end of a gesture and consecutive samples between start and end are extracted as a gesture. 

To minimize the probability of falsely extracting an un-occurred gesture, we further apply a gesture length constraint based on our experimental data. Figure~\ref{fig:plotLengthCDF} presents the CDFs of gesture durations when three subjects perform 6 different gestures. We can observe that around 90\% gestures are completed within 1s. Therefore, we apply a length constraint which ensures gestures less than 0.2s or greater than 1.4s are discarded. \textit{meanThr} and \textit{stdThr} are optimized through trial-and-error procedure and the values used in our work is 0.5 and 0.25 respectively. 

Figure~\ref{fig:SegmentationComparison1} illustrates the gesture segmentation result, where the green dots represent the start points and red squares represent the end points. Note that during a data collection session, the user always keeps his/her hand within the operating region thus avoiding any \textit{transition effects},  i.e., a slightly descending/ascending signal caused by entering/leaving the operating region. With the proposed segmentation algorithm, SolarGest successfully identifies 96\% of gestures in our dataset while incurring no false positives.

 \subsubsection{Signal Alignment}
 
 \label{s:alignment}

Using our simulator presented in Section \ref{sec:simulation}, we have identified specific alignment issues for gesture waveforms. We first explain these issues, followed by the techniques we have used to address them.

Figure~\ref{fig:simulParameters} studies the impact of 8 practical parameters, i.e., device parameters such as solar cell form factor and efficiency, environment parameters such as light intensity and size, as well as gesture parameters such as speed, hand angle, proximity, and displacement, on the gesture profiles. In each graph, only a specific parameter varies and the rest are set to default values as presented in Section~\ref{ss:estimating}. It can be observed that each parameter indeed affects the gesture waveform and the impact can be categorized into \textit{temporal variation} (variation in waveform duration) and \textit{amplitude variation}. Specifically, different gesture speeds and displacements lead to varying gesture durations, making the same gestures mismatched in time dimension. Other parameters, such as light intensity and hand angle (refers to the angle between hand and horizontal plane, as shown in Figure~\ref{fig:vertical}), result in amplitude shifts.

We apply Z-score transformation to align gesture amplitudes and dynamic time warping (DTW) to align gestures in time dimension. We illustrate the alignment process in Figure~\ref{fig:plotAlignment}, which plots two detected signals of \textit{LeftRight} gesture. We can see that there is an amplitude shift between the two signals as well as a mismatch between the peak-to-peak time difference. These mismatch effects may stem from variations in gesture proximity, light intensity, speed of hand motion, and so on. 
 
We first apply Z-score transformation, which is known to be an effective function to make multiple signal with different amplitudes comparable~\cite{cheadle2003analysis}. After transformation, distribution of the signal follows the normal distribution (i.e., mean 0 and standard deviation 1). Figure~\ref{fig:plotAlignment}(b) illustrates the waveforms after Z-score, in which we can see their amplitudes are converted to the same scale between [-2,2] and look very similar. After Z-score transformation, we can still observe the temporal misalignment issue. DTW has been successfully applied to various applications such as speech recognition~\cite{myers1980performance} and activity recognition~\cite{sempena2011human} to cope with such temporal mismatch. Thus, we apply DTW to gesture signals after Z-score transformation. The performance is shown in Figure~\ref{fig:plotAlignment}(c), from which we can see that the two signals almost overlap. With signal alignment, \systemName is able to minimize the impact of parameters that cannot be avoided in practical use due to human imperfection.

 \subsection{Feature Selection and Classification}

After signal processing, SolarGest extracts features from each detected gesture window and use them as the input for classification. In our system, two feature sets are considered and compared:
 
 \setlength{\leftmargini}{1.2em}
 \begin{itemize}
 \item \textbf{Statistical features:} include 22 typical time and frequency domain features, such as \textit{MEAN, STD, IQR, SKEW, KURT, Q2, DominantFrequency}, that have been widely used in human-related sensing~\cite{lara2013survey}. 
 
 \item \textbf{DWT coefficients: }  As mentioned before, DWT decomposes a signal and provides detail coefficients at multiple levels. Using these coefficients, DWT can perfectly reconstruct the original signal. Employing such DWT detail coefficients as features in classification has been extensively demonstrated in a wide range of human sensing applications~\cite{virmani2017position,abdelnasser2015wigest,wang2015understanding,tzanetakis2001audio}. Motivated by these applications, we extract DWT detail coefficients as another feature set. 
 \end{itemize}

Before applying DWT for feature extraction, we utilize spline interpolation to ensure that each detected gesture window has the same length of 512. The reasons for harmonizing the window lengths are three folds. First, classifiers require the same number of features for each gesture during training and classification. To obtain the same number of detail coefficients for each gesture, DWT requires detected gesture windows to have the identical length. Second, as DWT decimates the signal length by half at each level, it is efficiently calculated when the length of each gesture signal is a power of 2. We have measured the time duration of the six gestures and found that 90\% of them can be completed within 1s. Due to the sampling rate of 500Hz, we therefore interpolate each gesture to 512 samples. 

Then, we perform 5 level DWT decomposition on each gesture and extract the detail coefficients in the 5th level as the features. The reason to choose level 5 is explained in Section~\ref{s:denoising}. Regarding the selection of wavelet, five different wavelets are considered including: Haar(haar), Daubechies1(db1), Daubechies2(db2), Daubechies4(db4), and Coiflet2(coif2). Although Daubechies4 and Haar have shown good performance for Wi-Fi~\cite{virmani2017position} and light sensor based~\cite{venkatnarayan2018gesture} systems, we will investigate the effect of wavelet selection on solar-based gesture recognition in the evaluation section.

\begin{figure}[t]
    	\centering
    	\includegraphics[scale=0.92]{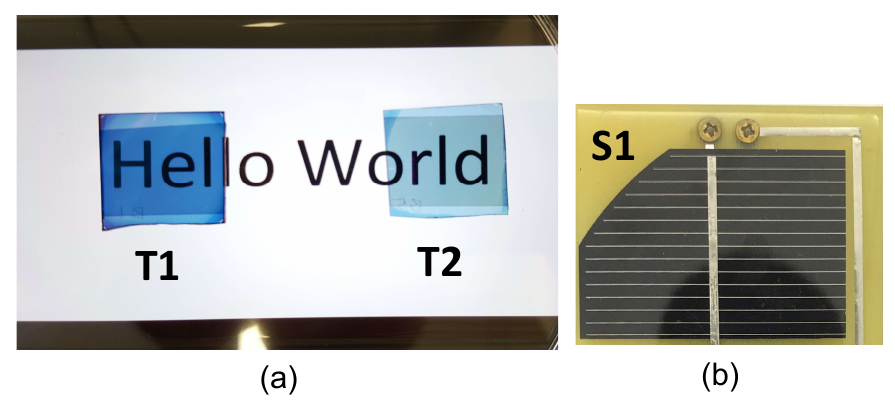}
    	\vspace{-0.15in}		
    	\caption{(a) Effect of placing the two transparent solar cells T1 and T2 on an iPhone 7 screen that displays the text `Hello Word'. (b) The silicon based solar cell S1.}
    	\label{fig:SolarCellPhone}
    	\vspace{-0.15in}
    \end{figure}
 
After feature extraction, machine learning classifiers are trained to recognize different gestures. In our system, we implemented four typical classifiers that are widely used for gesture recognition including: SVM, KNN, Decision Tree (DT), and Random Forest (RF). For DT, the confidence factor (C) and minimum number of instances (M) are set to 35\% and 2, respectively. For KNN, the number of nearest neighbors is set to 10 and the distance is weighted. For RF, the number of iterations (I) is set to 100. For SVM, we choose the cubic kernel. The performance comparison between different classifiers will be given in the following section.

 \section{Performance Evaluation}
 \label{sec:evaluation}
 
 We use real solar panels, both \textit{opaque} and \textit{transparent}, to evaluate our solar-based gesture recognition framework as well as qualitatively validate the simulation model derived in Section \ref{sec:simulation}.
 
 \subsection{Solar Cell Prototype}
 \label{ss:prototype}

As shown in Figure~\ref{fig:SolarCellPhone}, in our advanced photovoltaic laboratory, we prototyped three different solar cells, a 10x5cm silicon-based opaque solar cell (S1) and two 1x1cm transparent solar cells (referred to as T1 and T2) which were made from the same organic material (PBDB-T: ITIC) but with different transparencies (T1 20.2\% and T2 35.3\%) and thickness (T1 143nm and T2 53nm). To demonstrate the `see-through' property of the transparent solar cells, we placed them on the screen of an iPhone7. As shown in Figure~\ref{fig:SolarCellPhone}(a), we can clearly see the displayed `Hello World' context through both T1 and T2, but T2 provides a better `see-through' performance as it has a higher transparency to allow more visible light to pass through. In terms of the energy harvesting efficiency, T1 and T2 provide current densities of $13.82mA/cm^2$ and $6.85mA/cm^2$, respectively. More details of our transparent solar cell prototypes are available in~\cite{upama2017high1}. 

\begin{figure}[t]
      	\centering
      	\includegraphics[scale=0.4]{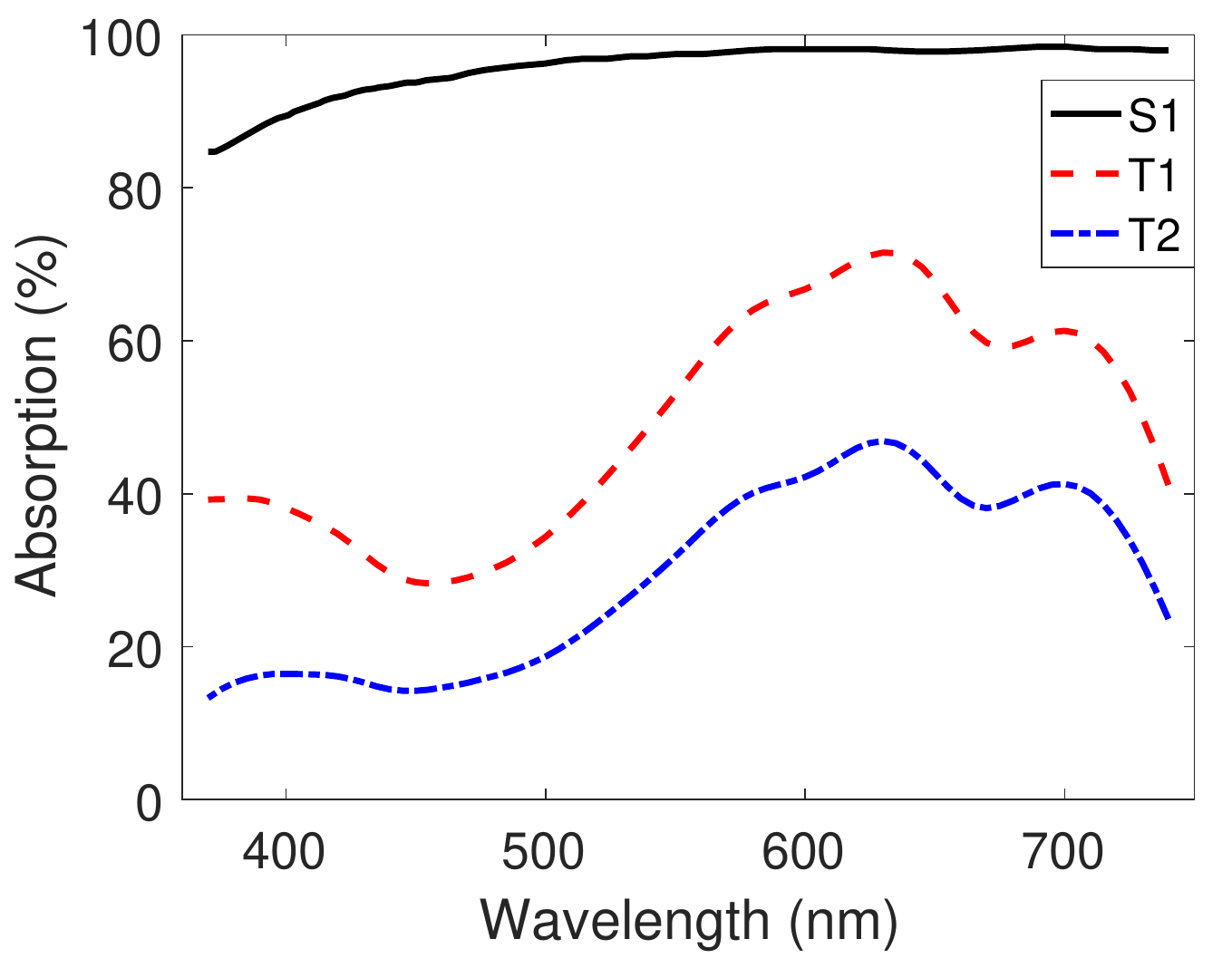}
      	\vspace{-0.1in}
      	\caption{Absorption spectra of the three solar cells S1, T1, and T2 within visible light band.}
      
      	\label{fig:plotAbsorption}
      	\vspace{-0.2in}
      \end{figure}

In Figure~\ref{fig:plotAbsorption}, we plot the absorption efficiency of the three solar cells in the visible light band. We can notice that the opaque solar cell S1 achieves nearly 100\% absorption efficiency over the entire wavelength range, whereas, the absorption rate of T1 and T2 is only 50\% and 30\%, respectively, on average. As discussed in our theoretical model in Section~\ref{ss:estimating} and will be verified in the following evaluation, the energy harvesting efficiency (i.e., the transparency) will affect recognition performance.

 \begin{figure*}[t]
             	\centering
             		\includegraphics[scale=0.76]{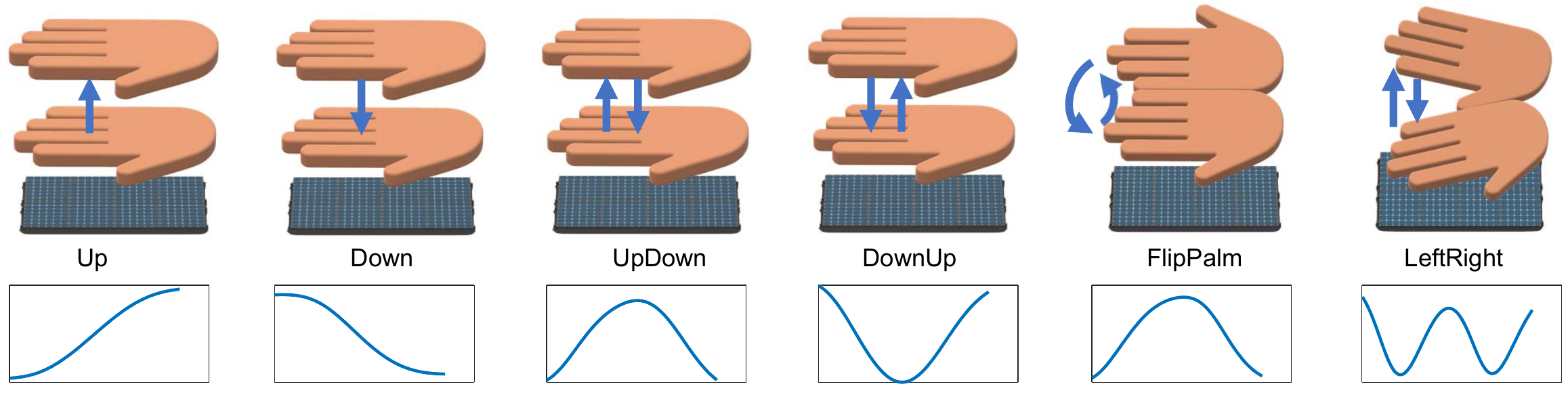}		
             	\caption{Illustrations of the 6 hand gestures conducted over the solar cells. The figures in the second row show the photocurrent profile collected under the 6 gestures.}
             	\label{fig:plotGestureIllustration}
 \end{figure*}
 
\begin{figure}[]
        	\centering
        	\includegraphics[scale=0.39]{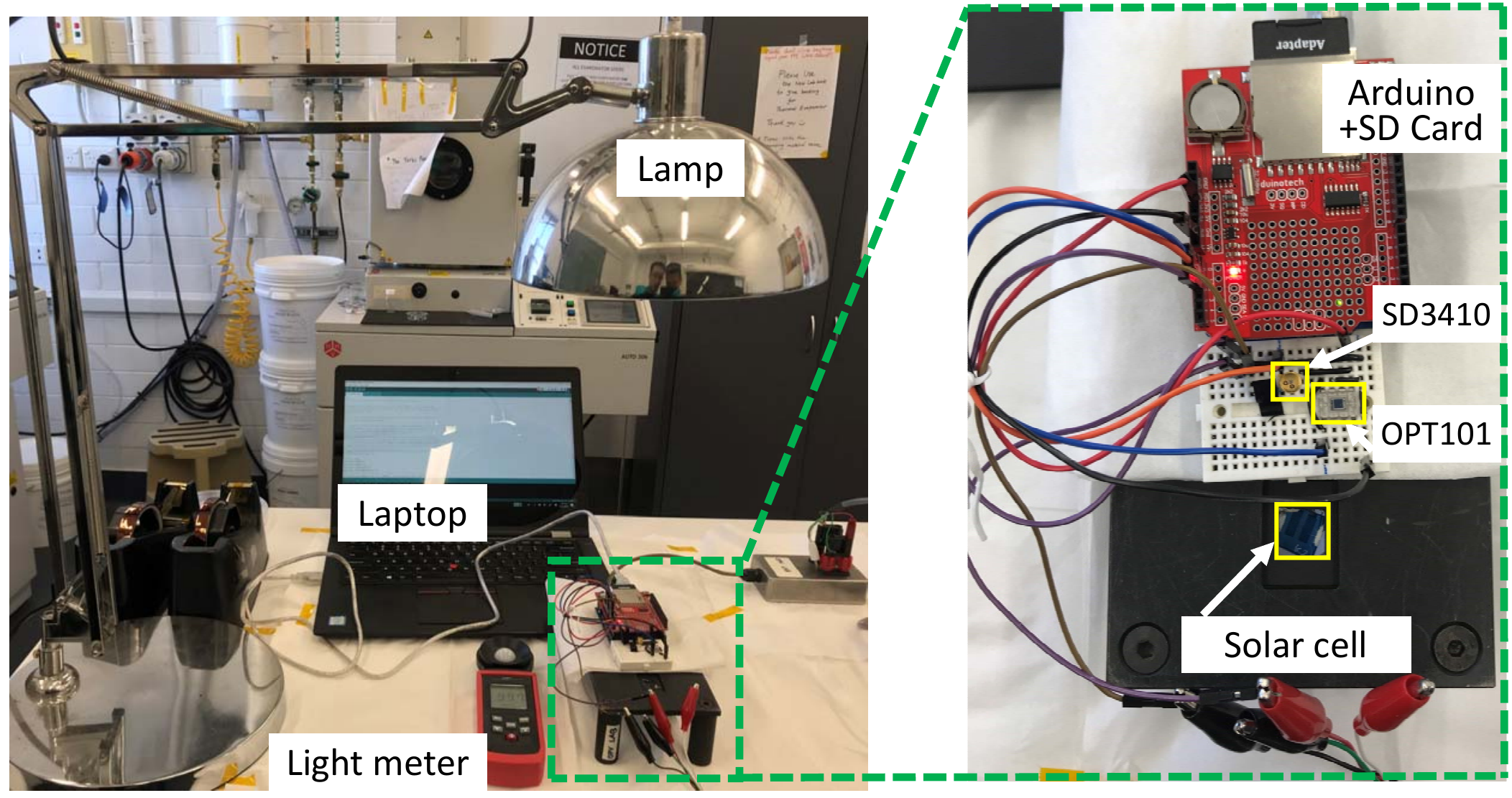}		
        	\caption{Data collection setup.}
        	\label{fig:dataCollectionSetup}
        	\vspace{-0.15in}
\end{figure}

 %\begin{figure}[t]
 %	\centering
 %	\subfigure[]{
 %		\includegraphics[scale=0.36]{fig/plotAbsorption.pdf}
 %		\label{fig:plotAbsorption}}
 %	\subfigure[]{
 %		\includegraphics[scale=0.65]{fig/FFT2.pdf}				
 %		%	\caption{FFT analysis of current signals collected from (a) indoor lighting environment and (b) outdoor lighting environment.}
 %		\label{fig:FFT}}		
 %	\caption{(a)  (b) FFT analysis of the signal collected from indoor and outdoor, respectively.}
 %\end{figure} 
 
 %Using the manufactured solar cells, we have collected real current signal when users are performing gestures. In detail, a user continually performs gestures among which a short pause (holding hand steady) is taken so that the exact gesture can be extracted by detecting its start and end. An Arduino Uno board is utilized to sample the current signal in 500Hz. Sampling in a high frequency also allows us to study the gesture recognition performance under different sampling rates by downsampling the signals.  

 \subsection{Gesture Data Collection}
 Using our prototype solar cells, we have collected a comprehensive gesture dataset for the performance evaluation of our proposed solar gesture recognition framework.\footnote{Ethical approval for carrying out this experiment has been granted by the corresponding organization.}. For data collection, we connect the solar cells to an Arduino Uno board as shown in Figure~\ref{fig:dataCollectionSetup}. The output of the solar cell is sampled by the Arduino via the onboard ADC at 500Hz and saved in the microSD card. For comparison purpose, we also collected the photocurrent signal from two different light sensors, TI OPT101 and Honeywell SD3410, which are widely used in ambient light based gesture recognition systems~\cite{li2016practical,li2017reconstructing,an2015visible,li2015human}. Figure~\ref{fig:dataCollectionSetup} illustrates our data collection setup at an indoor environment, which is conducted in our photovoltaic research lab due to the special (and bulky) tools required to connect the transparent cell output to the Arduino.

 During data collection, we have considered many different settings, including: (1) three solar cells with different energy harvesting efficiencies/transparencies; (2) five light intensity levels for indoor and outdoor combined (i.e., 800 lux and 2600 lux for transparent solar cells only under indoor lab environment; 10 lux, 50 lux, 800 lux, 2600 lux and 70k lux for the opaque solar cell under different scenarios including indoor and sunny outdoor); (3) six hand gestures as introduced in Figure~\ref{fig:plotGestureIllustration}; (4) threes subjects to perform the gestures; and (5) scenarios with/without human interference to investigate the robustness of \systemName against interference (data collected using the two transparent solar cells only). Specifically, the human interference is introduced by asking one subject to walk around in a half circle with radius of 30cm when another subject is performing gestures. As suggested by~\cite{li2018self}, light incident angles have little impact on the gesture recognition accuracy. Thus, we consider the case where light source is located at the top of the solar cell. 
 
 Table~\ref{tab:experiment setup} summarizes the considered experiment settings. In total, our data collection includes ten sessions (i.e., 2 transparent solar cell $\times$ 2 light intensities $\times$ 2 interference conditions + 1 opaque solar cell $\times$ 5 light intensities). In each of the session, subjects were asked to perform each of the 6 gestures 40 times. To avoid human fatigue, there was two minutes break between each session. The entire data was collected over five days. In total, we created a dataset consisting of $8\times3\times6\times40+5\times1\times6\times40$=6960 gestures.

\begin{table}[t]
\vspace{-0.2in}
           \small 
           \centering
           \setlength{\abovecaptionskip}{0pt}
           \setlength{\belowcaptionskip}{10pt}
           \caption{Experiment setting.}
           \label{tab:experiment setup}
           \ra{1.2}
           \begin{tabular}{@{}|l|l|l|@{}}
           \hline
           \textbf{Parameter}& \textbf{Option} & \textbf{Value}  \\ \hline
            \multirow{2}{*}{Solar cell} &\multirow{2}{*}{3} & transparent solar cell: T1, T2    \\
            & & opaque solar cell: S1  \\
             \hline
            Light intensity & 4 &10lux,50lux,800lux,2600lux,70klux \\ \hline
            Interference & 2 & with, without \\ \hline
            \multirow{2}{*}{Gesture} & \multirow{2}{*}{6} & Down, DownUp, FlipPalm, \\ 
            & & LeftRight, Up, UpDown \\ \hline
            Subject & 3 & 1 male, 2 female \\ \hline
            Photodiode & 2 &TI OPT101, Honeywell SD3410 \\ \hline
\end{tabular}
 \end{table} 

 \begin{figure*}[t]
   	\centering			
   	\includegraphics[scale=0.64]{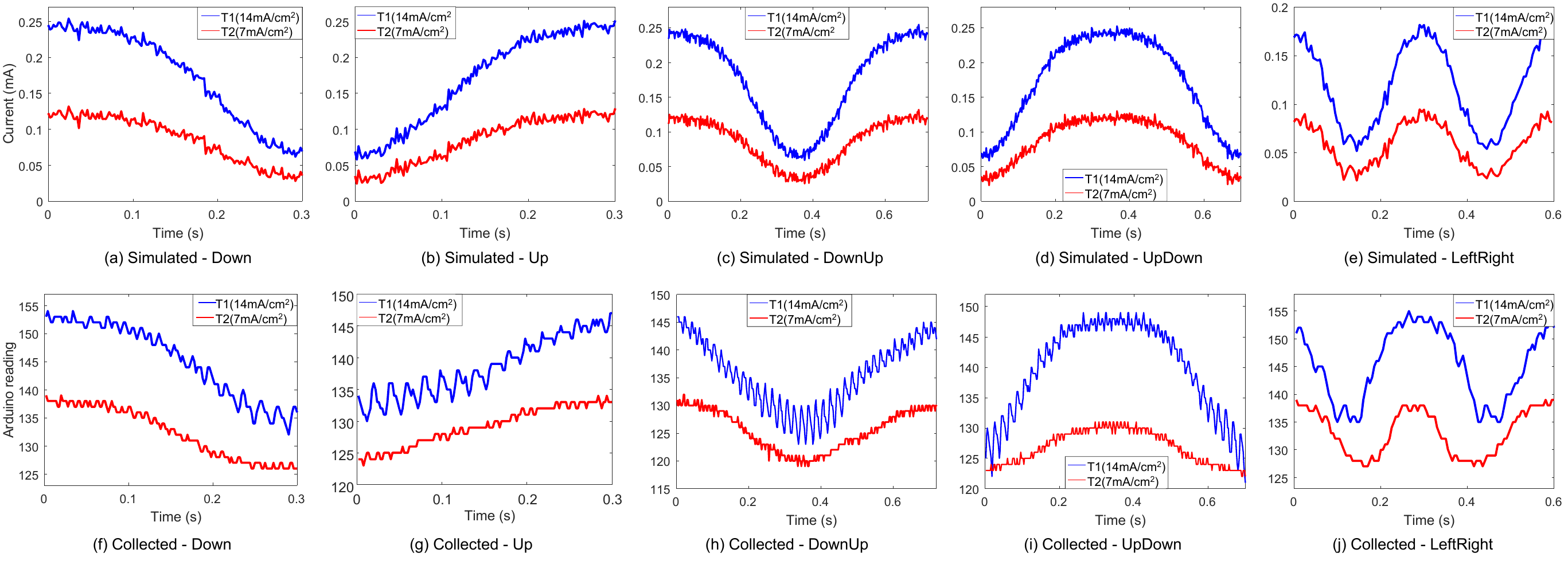}
   	\vspace{-0.2in}
   	\caption{Comparison of simulated gesture signal with the signal generated by solar cells.}
   	\label{fig:simulGestureCompare}
   	\vspace{-0.1in}
   \end{figure*}

 %We utilize a lamp to create 2 light intensities (i.e., 800 lux and 2600 lux), which is measured by a light meter (TASI 8121). We invited 3 subjects (including 1 female and 2 male) with age between 26 and 30 years old to collect data under all the scenarios. In terms of the outdoor experiment setup, we replace the transparent solar cell with the opaque one shown in Figure~\ref{fig:SolarCellPhone} (a). Only one subject is participated in outdoor experiment and human interference is not considered here.

  \subsection{Simulated vs. Real Waveforms}

 Figure~\ref{fig:simulGestureCompare} compares simulated gesture waveforms (top row) against actual waveforms (bottom row) collected from prototype transparent solar cells for 5 different gestures (note that FlipPalm is not captured in our model). It is clear that even though we model hand and solar cell as circles, the gesture signals simulated by our model are very similar to those generated by real solar cells in terms of signal features and patterns. This demonstrates that our model can be an effective tool to study gesture performance of next generation solar cells under a variety of scenarios.
 % that are either impossible or exhaustive in practice.

 \subsection{Gesture Recognition Performance}
  In this subsection, we evaluate the performance of our gesture recognition framework in terms of different system design choice and practical environmental factors. In addition, we also compare the performance of \systemName with that of light sensor based approaches. We use recognition accuracy, which represents the percentage of gestures that are correctly recognized by the classifier, as the metric. For each individual test, we perform 10-fold cross-validation and present the average recognition accuracy. The training and classification are implemented in Matlab.
 
% \begin{table}[]
% 	\centering
% 	\small
% 	\setlength{\abovecaptionskip}{0pt}
% 	\setlength{\belowcaptionskip}{10pt}
% 	\caption{Recognition accuracy given different transparencies and classifiers.}
% 	\label{tab:classifier}
% 	%\ra{1.2}
% 	\resizebox{2.in}{!}{
% 	\begin{tabular}{@{}ccccccc@{}}
% 		\toprule
% 		\multirow{2}{*}{\textbf{Classifier}} & \multicolumn{5}{c}{\textbf{Solar Cell}} \\ \cmidrule{2-6}
% 		&& \textbf{T1} & \textbf{T2}&& \\ \cmidrule{1-6}
% 		KNN &&97.9\% &  97.6\%&& \\ \cmidrule{1-6}
% 		Decision Tree && 96.3\%&95.1\% && \\ \cmidrule{1-6}
% 		SVM && 97.0\% & 96.2\% && \\ \cmidrule{1-6}
% 		Random Forest &&97.8\% &97.3\% && \\ 
% 		\bottomrule 
% 	\end{tabular}}
% 	\vspace{-0.1in}
% \end{table} 

  \subsubsection*{Comment 22:Sampling Rate} 
\begin{figure*}[]
    	\centering
    	\subfigure[]{
    	 		\includegraphics[scale=0.4]{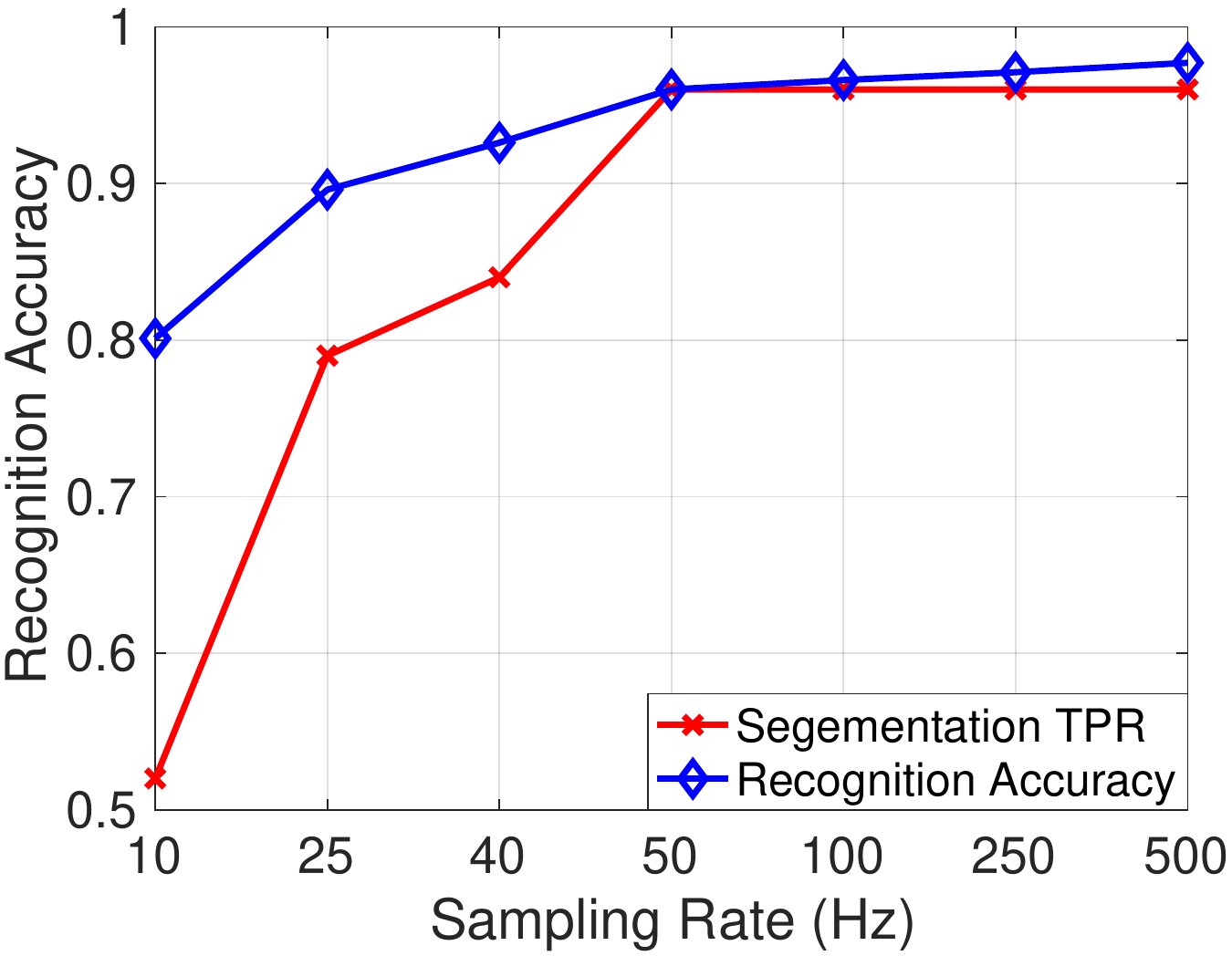}	
    	 		\label{fig:plotSamplingRate}}
    	\subfigure[]{
    		\includegraphics[width=5.6cm]{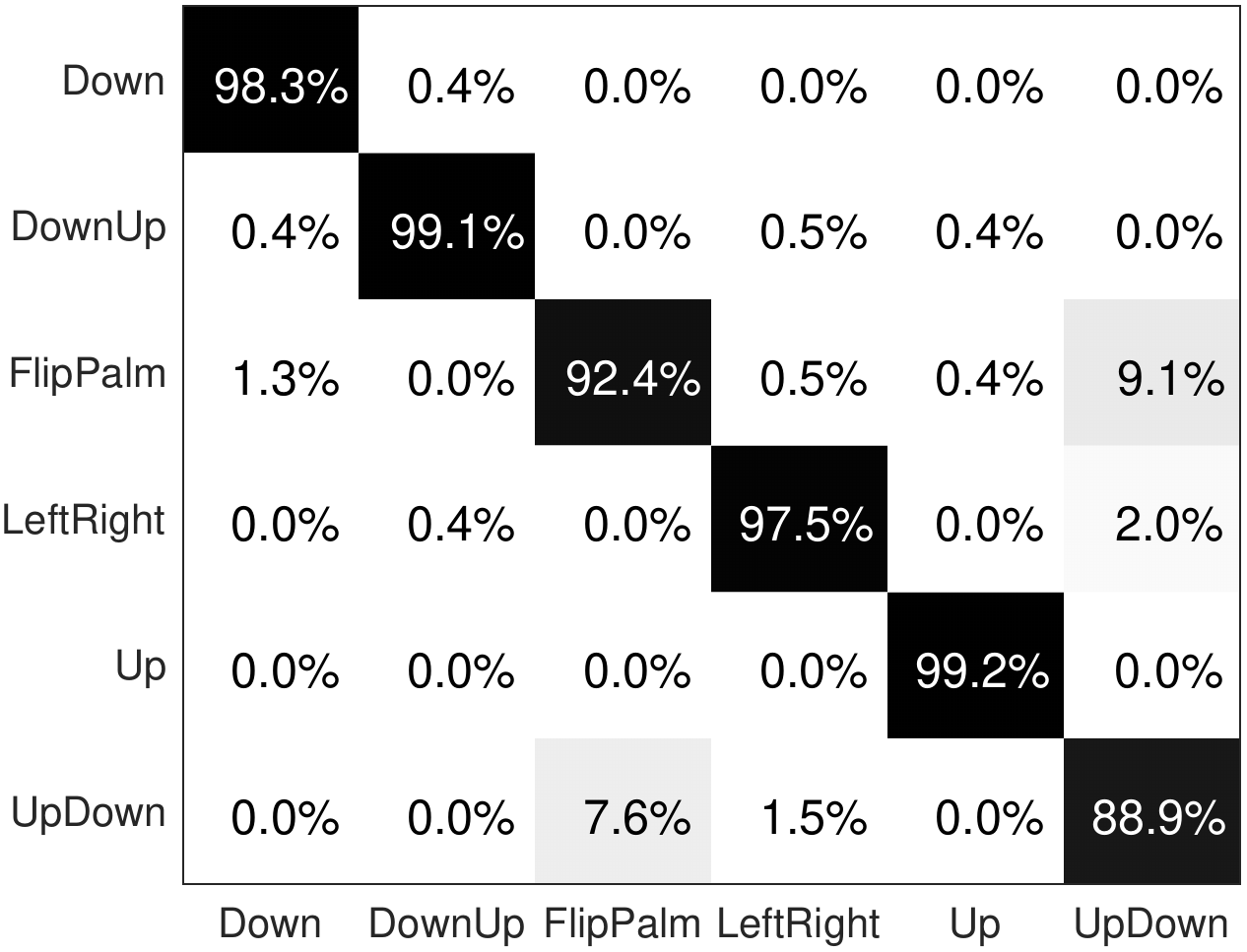}	
    		\label{fig:plotConfusionMatrixK2}}
    	%	\subfigure[]{
    	%		\includegraphics[scale=0.38]{fig/plotSamplingRate.pdf}	
    	%		\label{fig:plotSamplingRate}}
    	\subfigure[]{
    		\includegraphics[width=5.85cm]{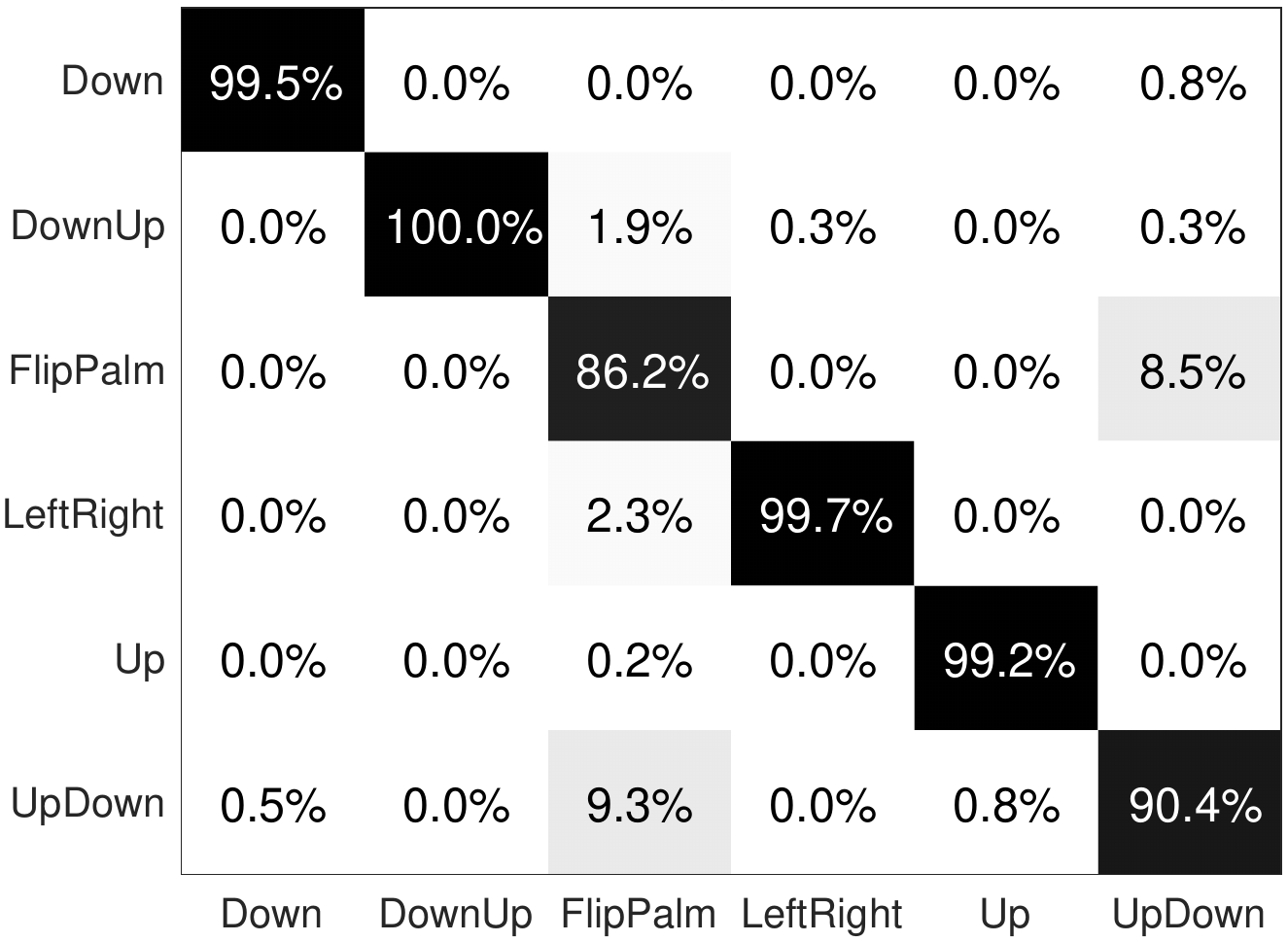}	
    		\label{fig:plotConfusionMatrixK7}}
    		\vspace{-0.2in}
    	\caption{(a) Recognition accuracy given different sampling rates; Confusion matrix of transparent solar cell (b) T1 and (c) T2.}
    	\vspace{-0.15in}
    \end{figure*}

 First, we investigate the minimum required sampling rate for SolarGest, as it directly affects the system power consumption. To achieve this, we down-sample the original 500Hz data to different sampling rates. Then, we apply the entire signal precessing and gesture recognition pipeline to each sampling rate. From Figure~\ref{fig:plotSamplingRate} we can observe that although  both segmentation and gesture recognition accuracies improve with increasing sampling rate at the beginning, performance stabilizes at 50 Hz. Therefore, we will consider a sampling rate of 50 Hz in the subsequent analyses.

 \begin{table}[t]
 	\centering
% 	\small
 	\setlength{\abovecaptionskip}{0pt}
 	\setlength{\belowcaptionskip}{10pt}
 	\caption{Recognition accuracy given different transparencies and classifiers.}
 	\label{tab:classifier}
 	%\ra{1.2}
 	\resizebox{3.2in}{!}{
 	\begin{tabular}{@{}ccccc@{}}
 		\toprule
 		\multirow{2}{*}{\textbf{Solar Cell}} & \multicolumn{4}{c}{\textbf{Classifier}} \\ \cmidrule{2-5}
 		&\textbf{KNN}  & \textbf{Decision Tree} & \textbf{SVM}&\textbf{Random Forest} \\ \cmidrule{1-5}
 		 \textbf{T1}&96.1\% & 94.1\%& 95.1\% &95.9\% \\ \cmidrule{1-5}
 		 \textbf{T2}&95.6\% & 93.0\%& 94.5\% &95.2\% \\ 
 	
 		\bottomrule 
 	\end{tabular}}
 	\vspace{-0.1in}
 \end{table} 
 
 \begin{figure*}[t]
   	\centering
   	\subfigure[]{
   		\centering
   		\includegraphics[scale=0.42]{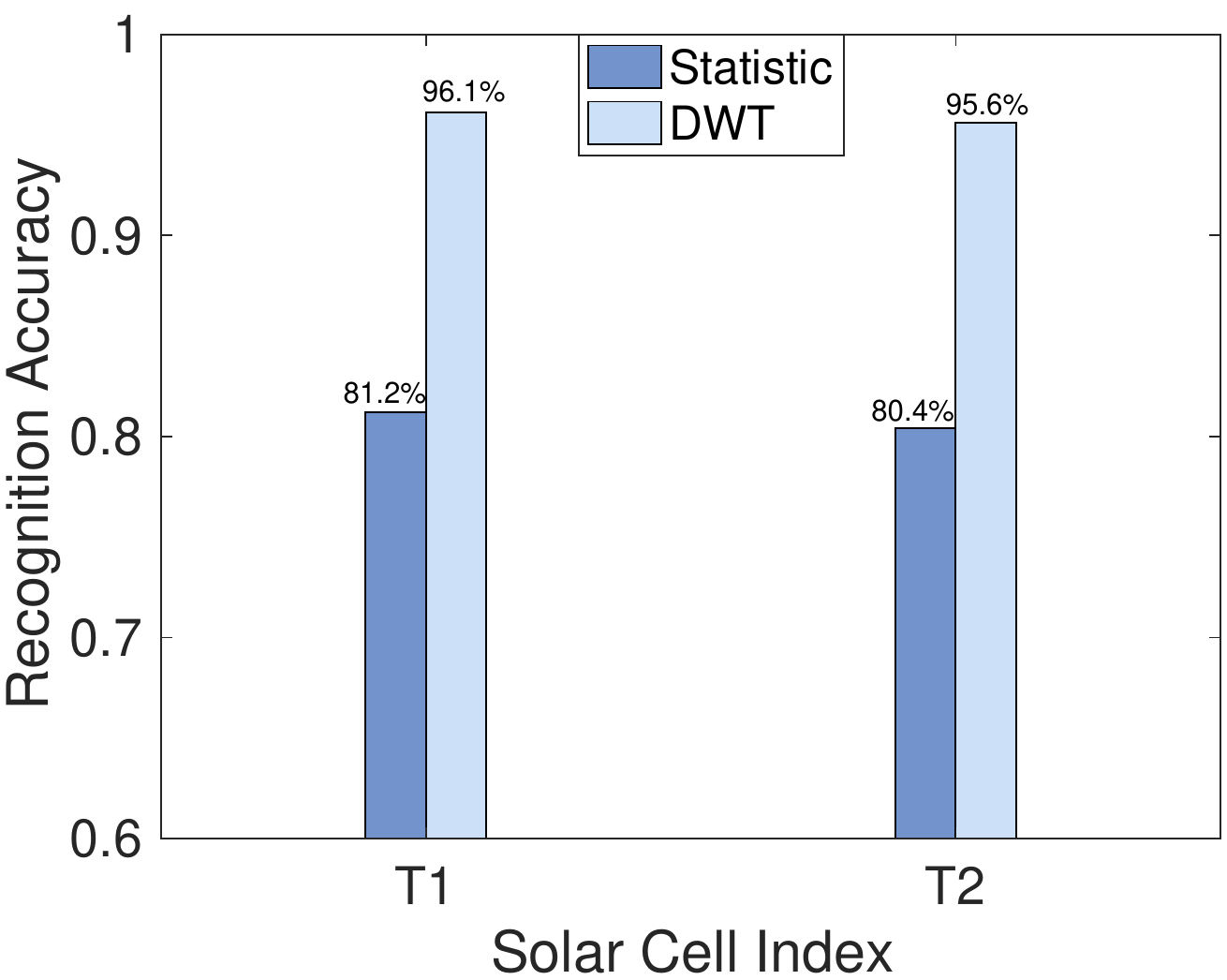}
   		\label{fig:plotFeatureSets}}
   	\subfigure[]{
   		\includegraphics[scale=0.42]{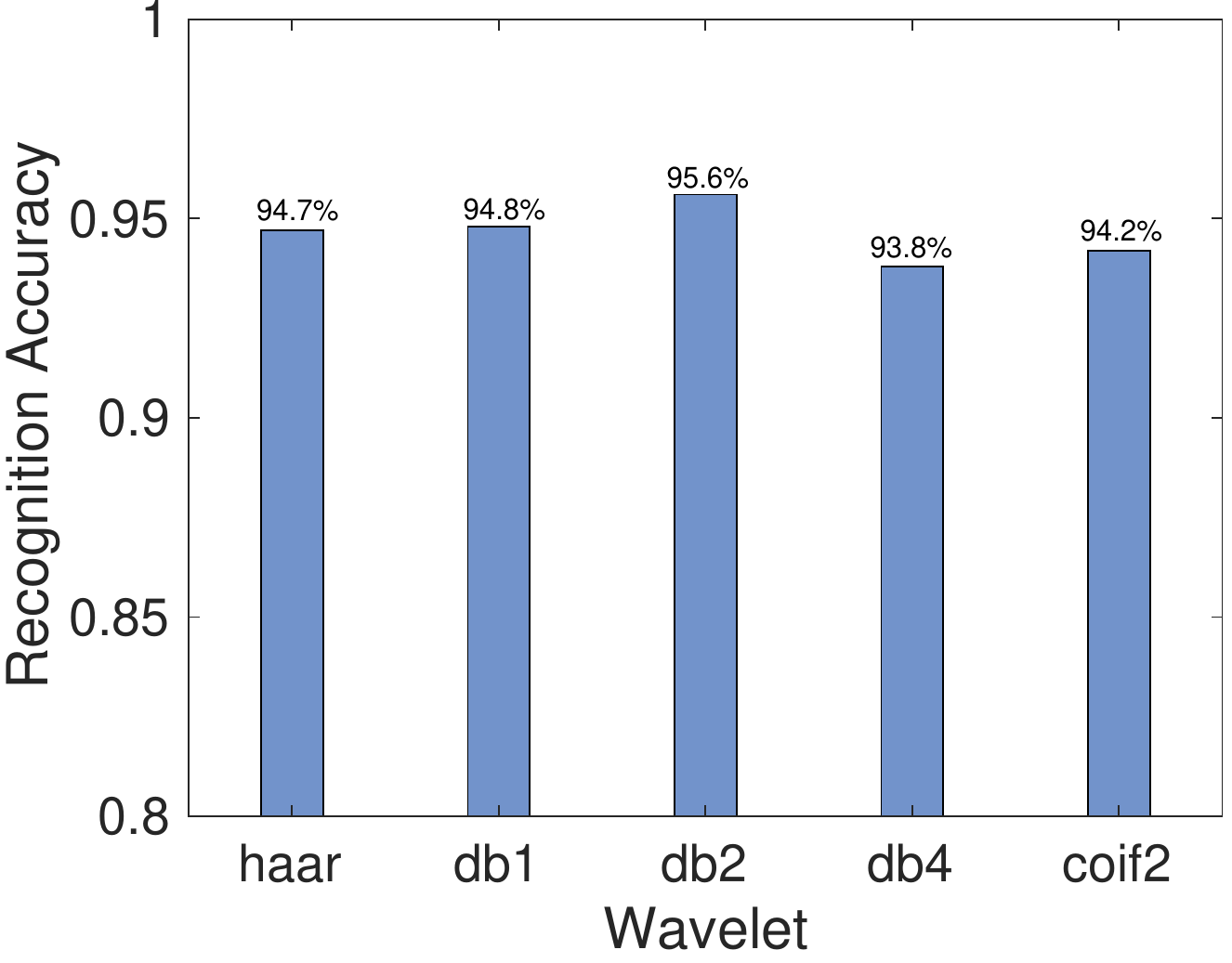}	
   		\label{fig:plotWavelet}}
   	\subfigure[]{
   		\includegraphics[scale=0.42]{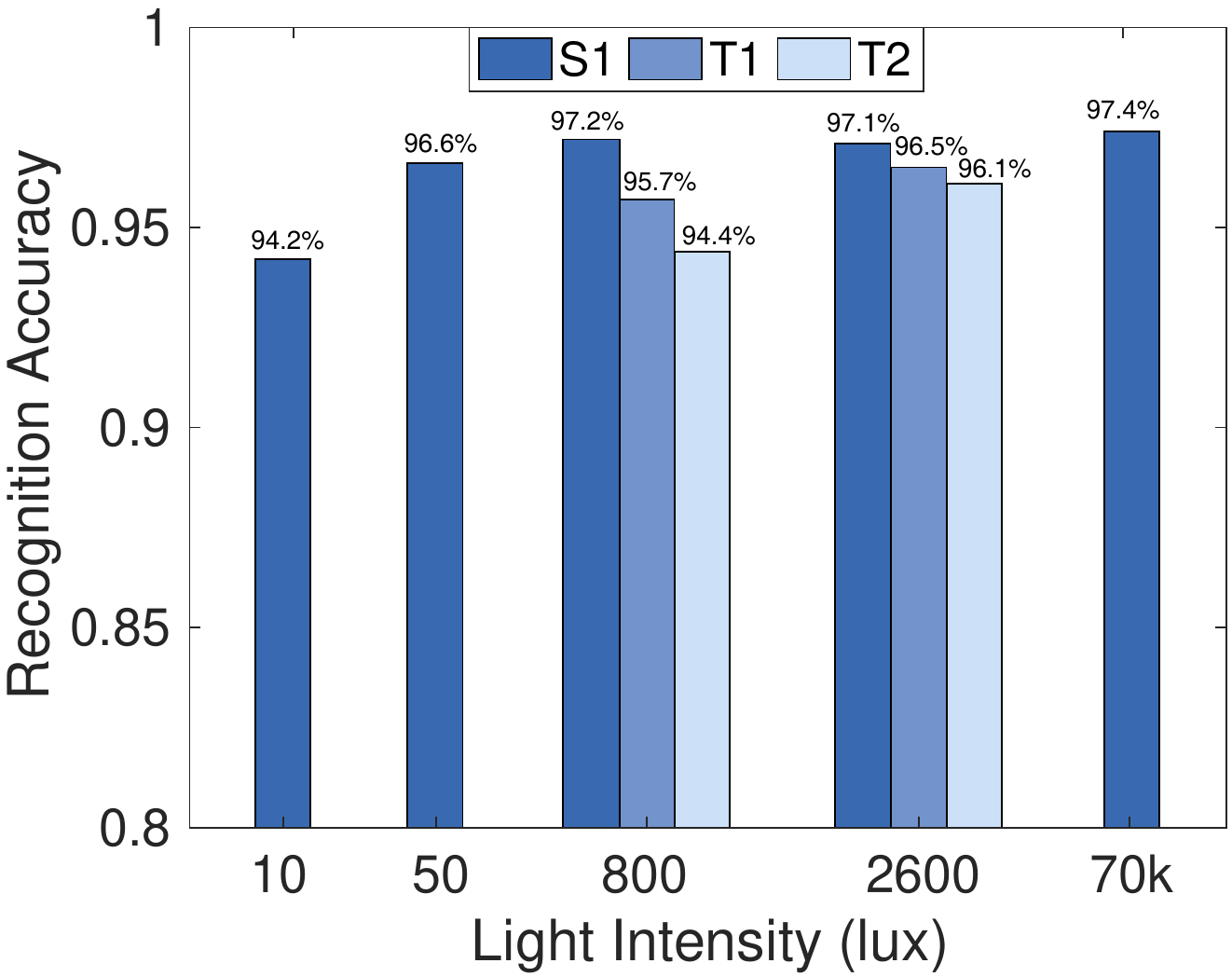}	
   		\label{fig:plotLightIntensity}}
   		\vspace{-0.2in}
   	\caption{Recognition accuracy given (a) different wavelets; (b) different feature sets; and (c) light intensity.}
   \end{figure*}
  \subsubsection*{Performance of transparent solar cells:}
 % In the following, we first investigate effect of different system design choices on the recognition performance, including: (1) the different solar cell transparencies; (2) selection of classifiers; (3) selection of feature sets; and (4) selection of DWT wavelet.
  
 Using DWT coefficients as features, Table~\ref{tab:classifier} presents average classification accuracies of the two transparent solar cells for the dataset obtained from all three subjects under 800 lux and 2600 lux indoor lighting with and without human interference. We can see that, despite being transparent with limited energy harvesting capacities compared to existing opaque cells, both T1 and T2 prototypes achieved very high accuracies under all four typical classifiers. This finding directly validates our earlier simulation-based predictions in Section \ref{ss:estimating}, which indicated that transparency will not reduce gesture recognition capability of solar cells in environments illuminated above 400 lux. Figure~\ref{fig:plotConfusionMatrixK2} - (c) show the confusion matrices of solar cell T1 and T2, respectively, which suggest that some gestures, e.g., FlipPalm and UpDown, are still likely to be confused with others as their patterns look very similar.
 
% \begin{figure}[t]
%   	\centering
%   	\includegraphics[scale=0.48]{fig/plotConfusionMatrixK2.pdf}		
%   	\caption{Confusion matrix of solar cell T1. }
%   	\label{fig:plotConfusionMatrixT1}
%   	\vspace{-0.1in}
%   \end{figure}

 \subsubsection*{Effect of Features:}
 Figure~\ref{fig:plotFeatureSets} compares accuracies when KNN is trained with different feature sets. As shown, DWT wavelets achieve approximately 98\% of accuracy compares to that of only 87\% for the statistical feature set. A more detailed wavelet analysis in Figure~\ref{fig:plotWavelet} reveals that Daubechies2 gives the highest accuracy for our solar cell based gesture recognition system, although Haar was reported to be the best wavelet for light sensor based gesture recognition \cite{venkatnarayan2018gesture}.
 
  \subsubsection*{Effect of environment factors:}
 % In the following, we first investigate effect of environment factors on the recognition performance, including: different light intensities; the impact of ambient user interference, and the recognition accuracy in unseen scenarios. 
  
 %Light intensity is an important parameter in practical use of \systemName as people stay in different environments (e.g., indoor and outdoor, dark and bright) with the light intensity significantly varies. 
 
 First, we test five light intensity levels that correspond to common conditions, 10 lux-dark room, 50 lux-living room, 800 lux-office, 2600 lux-cloudy, and 70k lux-sunny. Transparent solar cells are tested under 800 lux and 2600 lux only due to the sensitivity to environment (e.g., humidity), while the opaque solar cell is tested under all the five conditions. Figure~\ref{fig:plotLightIntensity} presents the recognition accuracy of the three solar cells under the five intensity levels. We can observe that, for the same solar cell, higher light intensity ensures a higher recognition accuracy, although the improvement is minor. This indicates that for common environment, \systemName is able to achieve superior and consistent performance. We can notice that S1 obtains higher accuracy compared to T1 and T2 at 800 lux and 2600 lux. The reason is that the energy harvesting efficiency as well as the form factor of S1 is larger than that of T1 and T2.

To assess the limit of SolarGest, we create an extremely dark environment (i.e., 10 lux) by turning off all lights in a dark room except a laptop screen. We find that, with our current prototype (Arduino UNO), the collected signal always remains zero, making it impossible to detect any gesture. The reason is that the resolution of Arduino ADC (10bit) is not enough to capture the minor changes in photocurrent. However, we found that dark environment problem can be solved by using either a high resolution ADC (e.g., 16bit) or amplifying the current. We implemented an amplification circuit and tested two amplification factors: 32$\times$ and 64$\times$. The results show that, with both amplification levels, gesture accuracy reaches to around 94\%.
 
\begin{figure}[t]
    	\centering
    	\subfigure[]{
    		\includegraphics[scale=0.345]{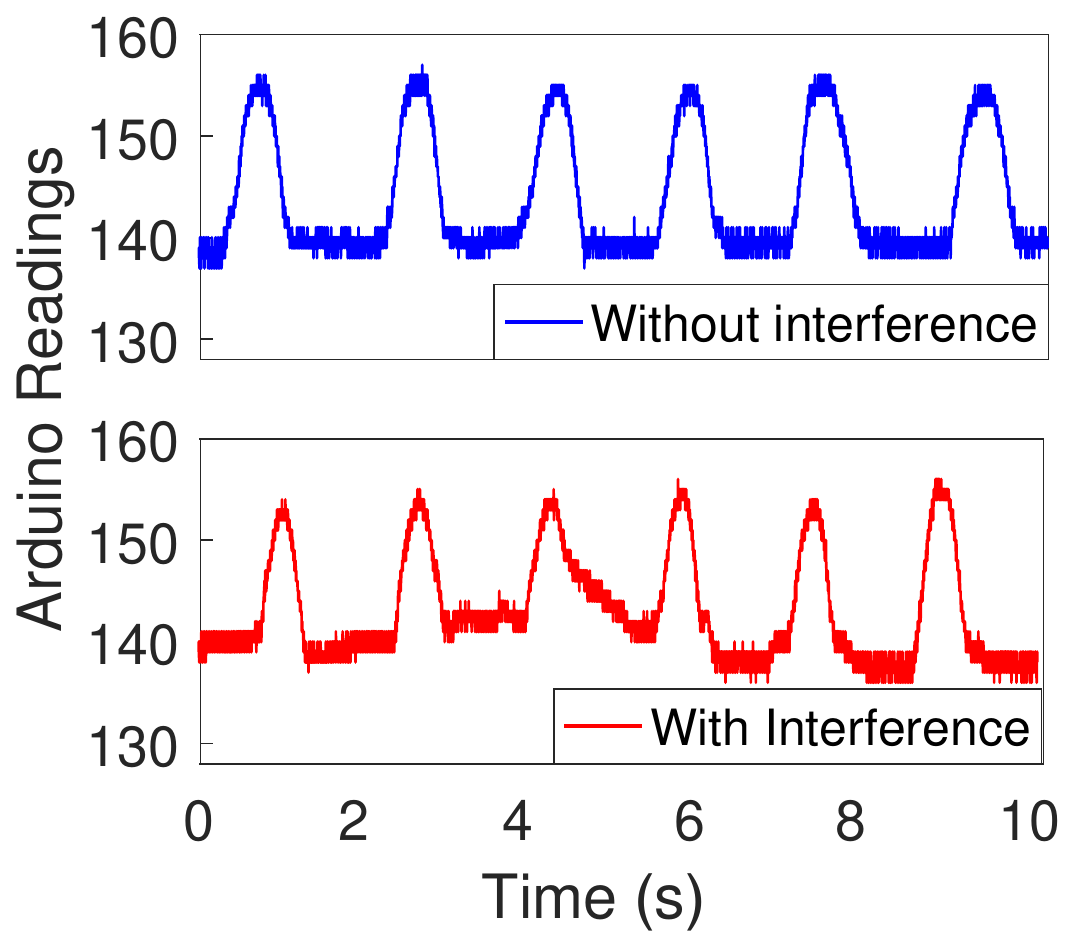}
    		\label{fig:plotInterferenceRawSignal}}
    	\subfigure[]{
    		\includegraphics[scale=0.315]{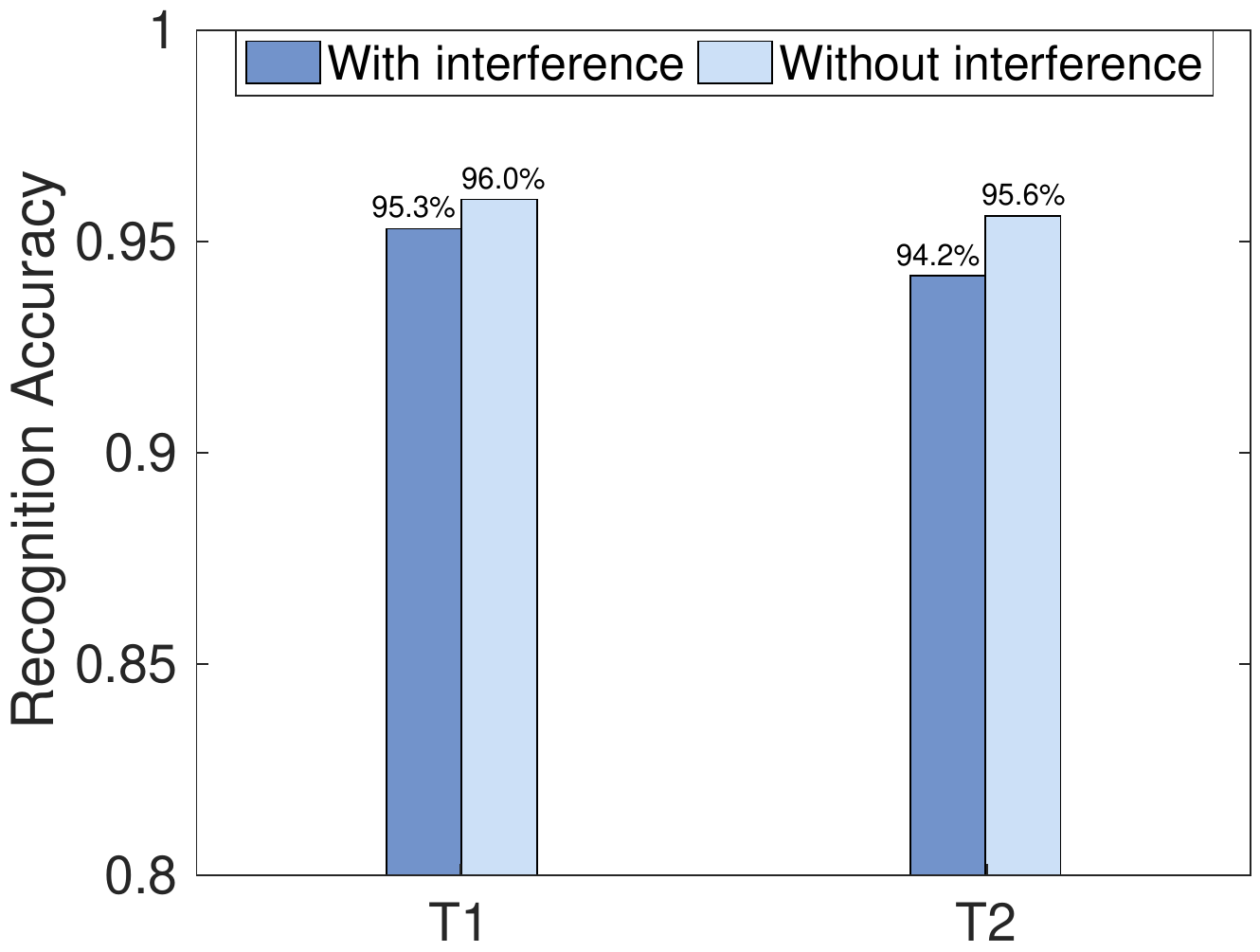}	
    		%\caption{Impact of interference on recognition accuracy.}
    		\label{fig:plotInterference}}	
    		\vspace{-0.2in}
    	\caption{(a) Comparison of raw signals with and without interference. (b) Impact of interference on recognition accuracy.}
    \vspace{-0.1in}
    \end{figure}

Second, we investigate the robustness of \systemName against ambient human interference in Figure~\ref{fig:plotInterferenceRawSignal}. We can see that human walking near the solar panel introduces some fluctuations in the signals. To investigate the impact of such signal interference on gesture accuracy, Figure~\ref{fig:plotInterference} plots accuracy with without human interference, which shows that interference reduced accuracy by only 1.5\% and \systemName still achieved 96\% recognition accuracy. 
 
Third, we investigate the gesture recognition performance when subjected global light intensity changes (e.g., walking from indoor to outdoor or sunlight is blocked by cloud during a gesture). We conduct the experiment using the simulator presented in Section 2. Specifically, we train the model using gestures simulated under stable light intensity, while test using the distorted gestures (simulated by switching the light intensity in different levels and frequencies) only. Our results indicate that when light intensity changes very fast (e.g., >50Hz), the accuracy is not affected, while almost half of the distorted gestures are wrongly recognized when intensity switching rate is low (e.g., 2Hz). However, as suggested in~\cite{li2018self}, such low dynamic global light change can be effectively filtered out by subtracting it.

\subsubsection*{Performance under Unseen Scenarios}
\begin{figure}[t]
      	\centering
      	\subfigure[]{
      		\includegraphics[scale = 0.295]{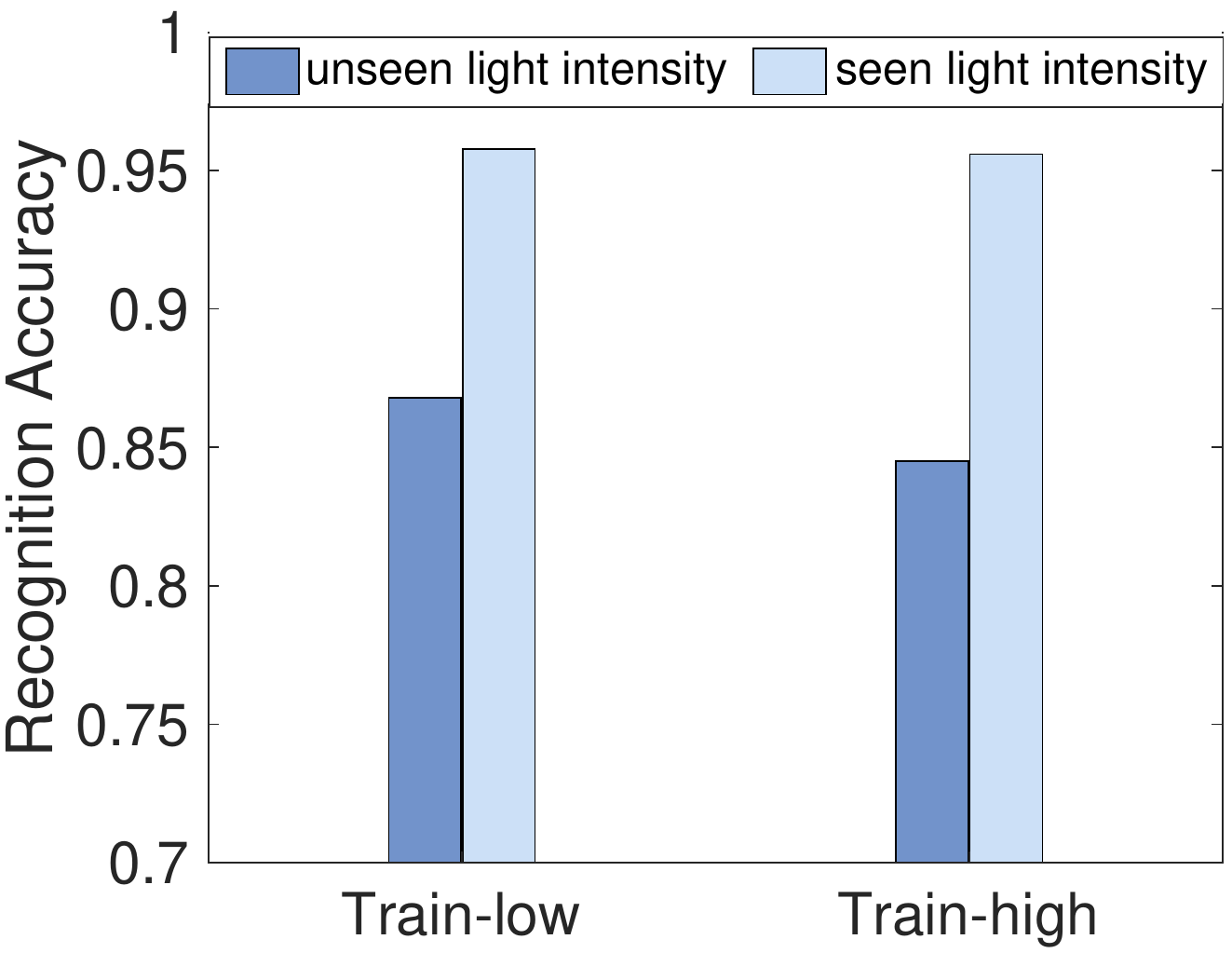}	
      		\label{fig:plotLightDependency}}
      	\subfigure[]{
      		\includegraphics[scale=0.295]{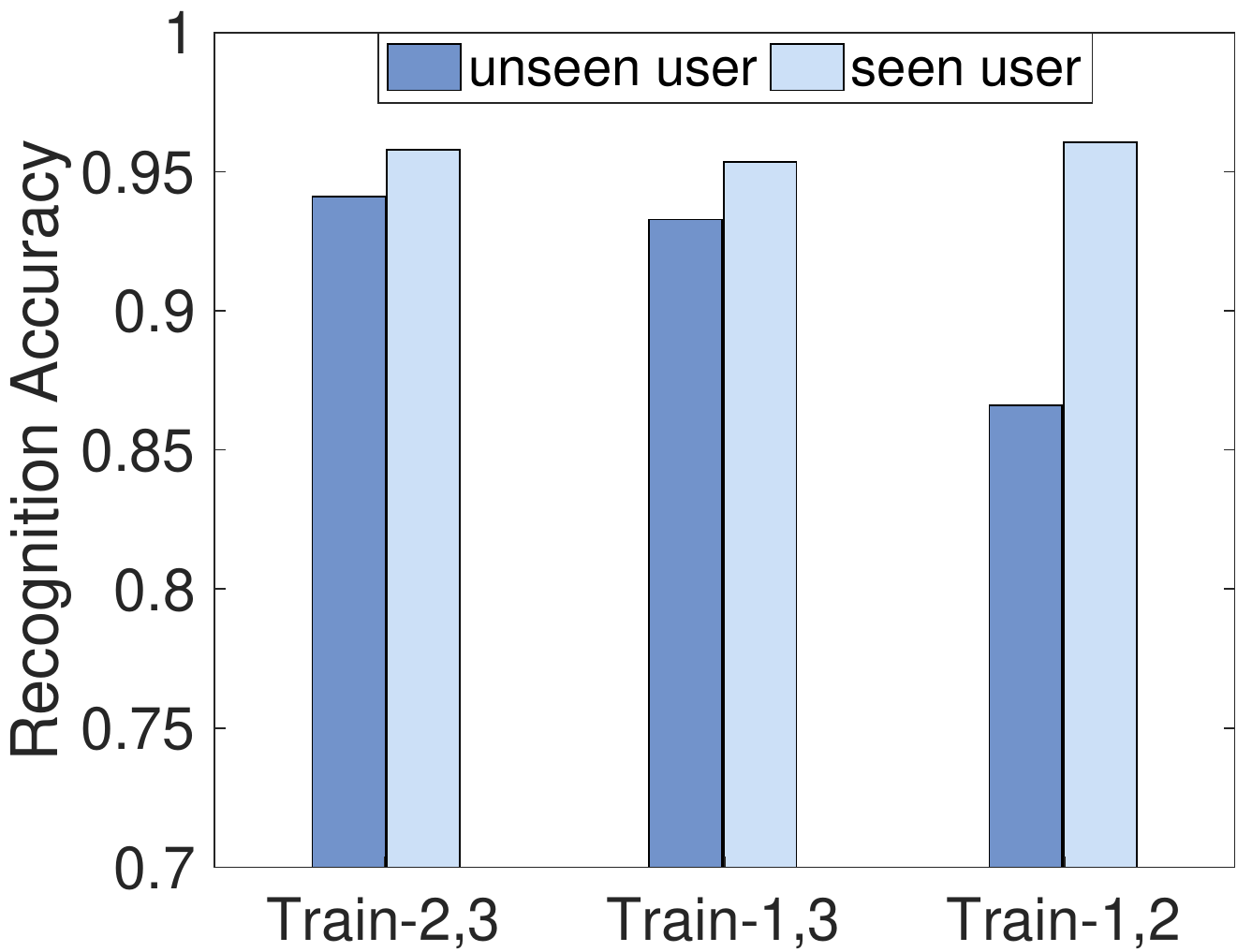}	
      		\label{fig:plotUserDependency}}
      		\vspace{-0.2in}
      	\caption{Recognition accuracy on (a) unseen lighting environment, (b) unseen user.}
      	%\label{fig:plotLengthCDF}
      	\vspace{-0.2in}
      \end{figure}

 We consider two unseen cases. First, we train the classifier using the data collected under one light intensity and test it by the data collected under another light intensity. The performance of training and testing with the same light intensity, i.e., seen scenario, is also obtained. From Figure~\ref{fig:plotLightDependency}, we can see that \systemName still achieves 88\% accuracy even in unseen lightning environment case. Second, we train the classifier using the data collected from two subjects and test it on the remaining one. The results in Figure~\ref{fig:plotUserDependency} indicate that \systemName is robust to subject difference. Although the accuracy suffers from a significant drop when training with subject1 and subject2, but testing with subject3, \systemName achieve 93\% accuracy on average for unseen users. As a result, training on a large number of subjects may not be necessary.

\subsubsection*{Comparison with light sensor based systems}

%\sout{Earlier in Figure~\ref{fig:SegmentationComparison}, we observed that signal traces from light sensors are noisier compared to those from the solar cells, which affect gesture segmentation performance. For instance, as shown in Figure~\ref{fig:SegmentationComparison}, by using the signals from solar cell, the system can perfectly detect all the ten gestures, whereas, both of the two light sensors can only detect eight of the ten gestures.}  

\begin{figure}[t]
       	\centering
       	\includegraphics[scale = 0.57]{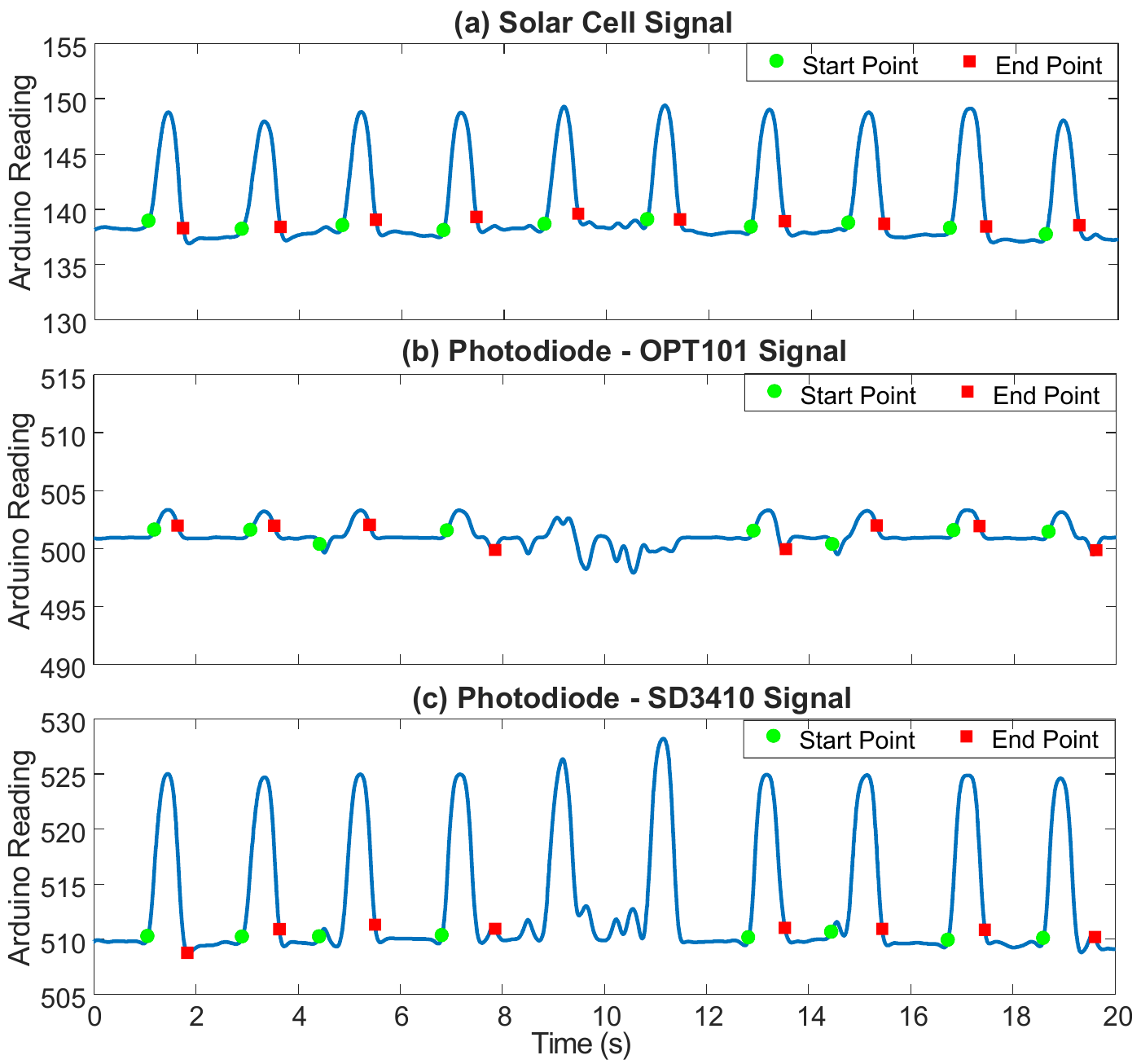}
       	\caption{Segmentation performance comparison using signals from (a) solar cell T1, (b) photodiode OPT101, and (c) SD3410, under gesture FlipPalm. The green dots represent the detected start points and the red squares represent the detected end points.}
       	\vspace{-0.2in}
       	\label{fig:SegmentationComparison}
       \end{figure}

Figure~\ref{fig:SegmentationComparison} compares the signal traces from solar cell T1 and two light sensor collected at the same time, where we can observe that signal traces from light sensors are noisier. As a result, by using the signal from solar cell, the system can perfectly detect all the ten gestures, whereas, both of the two light sensors can only detect eight of the ten gestures. Table~\ref{tab:overall accuracy} compares the overall performance of light sensors and transparent solar cells in terms of both segmentation and recognition accuracies. We can notice that solar cells achieve 12\% to 26\% higher segmentation accuracies and at least 5\% better gesture recognition accuracy compared to light sensors. These results demonstrated that, for gesture recognition, even transparent solar cells are no worse off than light sensors.

 \section{Power Measurements}
 \label{s:power}
 In this section, we investigate the power saving advantage of \systemName against conventional light sensor based systems. As shown in Figure~\ref{fig:pipeline}, the power consumption of SolarGest comes from two parts: MCU sampling and data transmission. In contrast, light sensor based systems consumes additional energy in powering the light sensors. In the following, we perform a conservative comparison which assumes that only one sensor is required for light sensor based systems (current works require a array of sensors~\cite{kaholokula2016reusing,venkatnarayan2018gesture}).
 
 \begin{table}[t]
   	\centering
%   	\small
   	\setlength{\abovecaptionskip}{0pt}
   	\setlength{\belowcaptionskip}{10pt}
   	\caption{Performance comparison between light sensors and solar cells.}
%   	\vspace{-0.15in}
   	\label{tab:overall accuracy}
   	%\ra{1.2}
   	\resizebox{3.3in}{!}{
   	\begin{tabular}{@{}ccccc@{}}
   		\toprule
   		
   		\textbf{Metric} & \textbf{OPT101} 
   		& \textbf{SD3410} & \textbf{SolarGest(T1)}&\textbf{SolarGest(T2)} \\ \cmidrule{1-5}
   		
   		Segmentation Accuracy & 70.2\% & 84.3\% & 96.2\% & 96.1\% \\ \cmidrule{1-5}
   		% Segmentation FPR & 6.7\% & 1.4\% & 0.0\% & 0.3\% \\ \cmidrule{1-5}
   		Recognition Accuracy & 55.2\% & 89.9\% & 96.1\% & 95.6\% \\ 
   		
   		\bottomrule 
   	\end{tabular}}
   	\vspace{-0.2in}
   \end{table}   
 
 \textbf{MCU Power Measurement:} since both solar cell and light sensor are sampled by analog-to-digital converter (ADC), we conducted an experiment to measure the power consumption in ADC sampling. We select the Texas Instrument SensorTag as the target device, which is equipped with an ultra-low power ARM CortexM3 MCU. The SensorTag is running with the Contiki operating system. As discussed in Figure~\ref{fig:plotSamplingRate}, to achieve over 95\% of accuracy, a sampling rate of 50Hz is required by \systemName. Thus, we duty-cycled the MCU at 50Hz for sampling and applied an oscilloscope to measure the average power consumption of SensorTag during the sampling. According to our measurement, the system consumes 20.28$\mu$W in sampling the signal at 50Hz.

\begin{figure}[]
  	\centering
  		\includegraphics[scale=0.35]{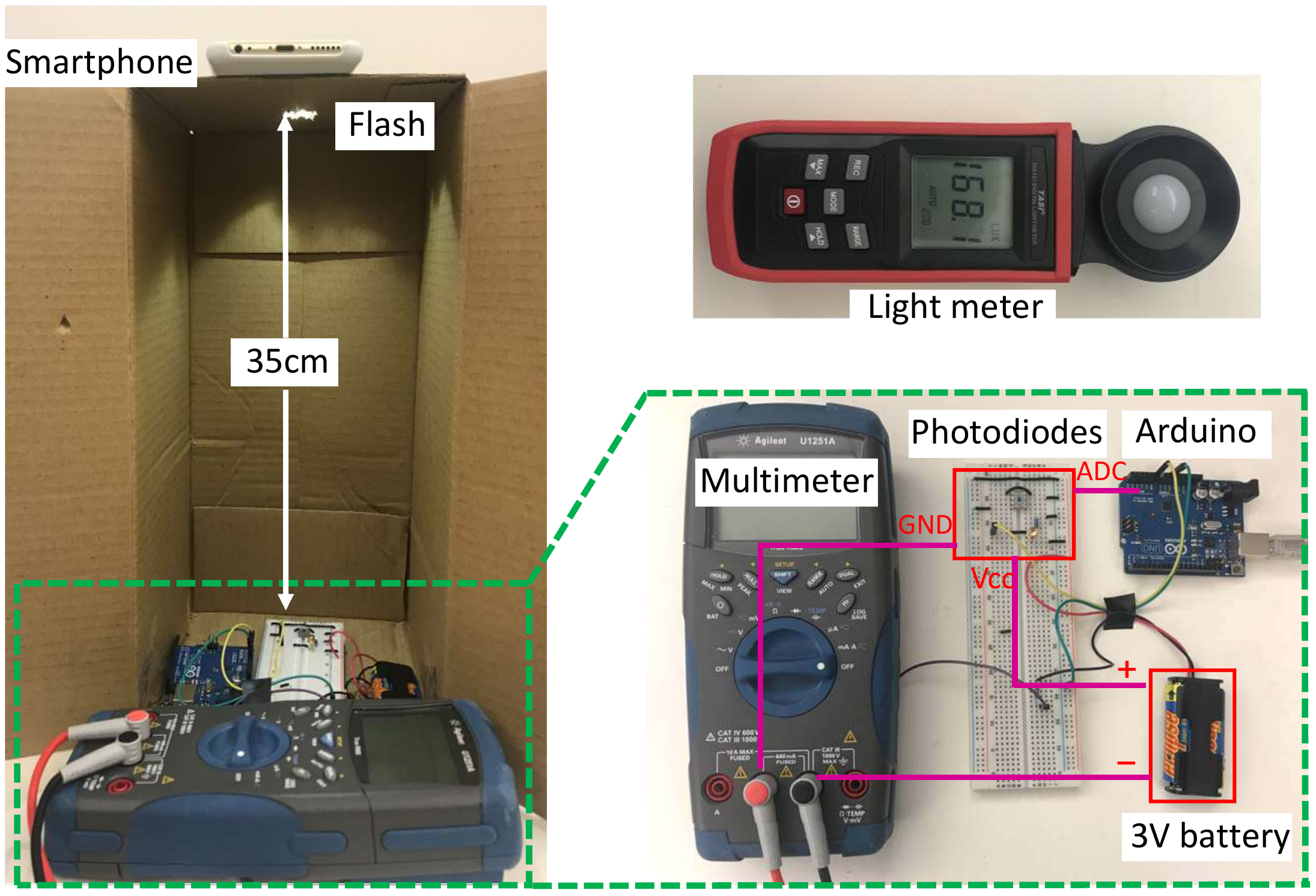}	
  	\caption{Light sensor power measurement setup.}
  	\label{fig:powerPDSetup}
  	\vspace{-0.15in}
  \end{figure}

\begin{figure}[]
   	\centering
   	\includegraphics[scale=0.46]{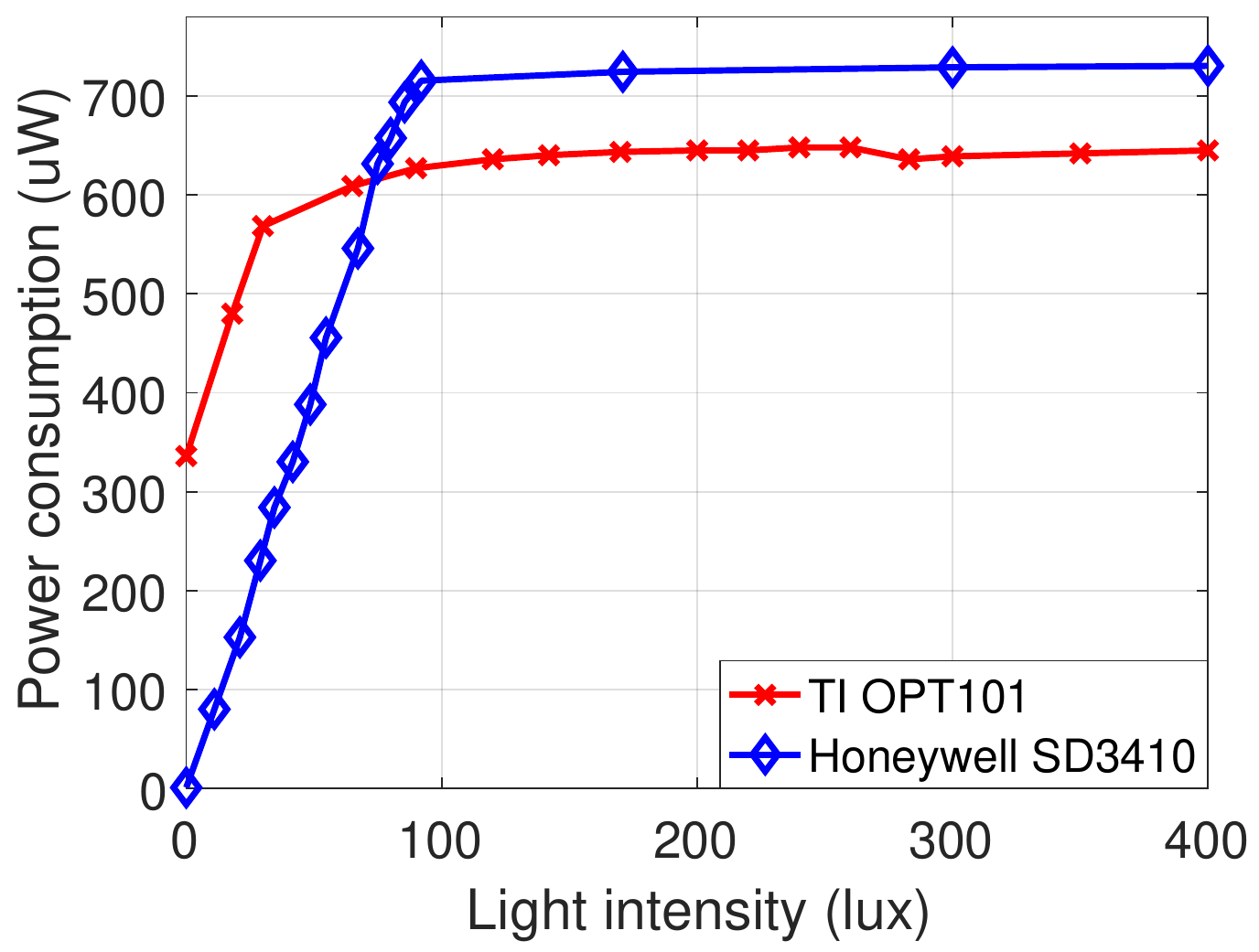}	
   	\caption{Power consumption of light sensors with different light intensities.}
   	\label{fig:plotPowerConsumption}
   	\vspace{-0.15in}
   \end{figure}

 \textbf{Light sensor Power Measurement:} in addition, we also measure the power consumed by the light sensor itself. We consider two light sensors, namely TI OPT 101 and Honeywell SD 3410, that are widely used in the literature~\cite{li2015human,an2015visible,li2017reconstructing,li2016practical}. In particular, we measured the power consumption of the sensors under different light intensities (assuming normal operation scenarios), as the datasheet only gives the power consumption when the sensor is operated in dark environment. Figure~\ref{fig:powerPDSetup} illustrates the measurement setup. To minimize the effect of ambient light, we conduct the experiment in a box with one side open. A smartphone is placed on top of the box and its Flash is used as the light source. We create an aperture with a radius of 1cm on the top of the box and place the light sensor right below the aperture to ensure $0^\circ$ of light incident angle. The light sensor is powered by a 3V battery and a multimeter is used to measure the current. Figure~\ref{fig:plotPowerConsumption} presents the power consumption of the two light sensors under different light intensities. We can observe that the power consumption is not constant. When the light intensity is lower than 100 lux, the power consumption increases linearly with light intensity. Once the light intensity is higher than 100 lux, the energy consumption becomes stable. Since the light intensity of normal environment is usually higher than 100 lux, e.g., 200-800 lux for office environment, it means that, without duty-cycling (sensor always turn on), OPT101 and SD3410 consumes around 650uW and 730uW, respectively. With 50 Hz duty-cycle, the power consumption reduces to 39.78 $\mu W$ and $42.18 \mu W$, respectively. In addition, our results is consistent with the datasheet when light intensity is 0~\cite{OPT101datesheet}. In contrast, solar cell is passive and does not require any external power.

 \textbf{Overall System Power Saving:} Now, we analyze the overall system power consumption. Considering 50Hz sampling rate and a duty-cycled system, Table~\ref{tab:power analysis} compares the power consumption of SolarGest and light sensor (i.e., photodiode) based system. Note that the photodiodes are assumed to operate in photoconductive mode, which requires external power supply, in order to provide faster response rate~\cite{chen2015high}. The recent advancement in Wi-Fi backscattering has demonstrated that 1 Mbps data rate can be achieved with only 14.5 $\mu W$ power consumption~\cite{kellogg2016passive}. Given a sampling frequency of 50Hz, \systemName has 100 Bytes data (2 Bytes for each 12-Bits ADC reading) to be transmitted per second. Thus, it means that 0.023 $\mu W$~\footnote{$P_{backscatter} = (100*8)/1000000*14.5 \mu W = 0.023\mu W.$} is required for backscattering-based data transmission. Overall, the power consumption of \systemName will be around 20.3 $\mu W$, while the consumption of light sensor based system is about 60.1 $\mu W$. Thus, \systemName is able to save over 66\% of the energy. In a more general case where BLE is used for communication, 31.11$\mu W$ power is required for the data transmission (100 Bytes per second) based on our measurement (using TI SensorTag as the target platform.). In this case, the overall system power consumption for \systemName and the two light sensors based system increase to 51.39 $\mu W$, 91.17 $\mu W$, and 93.57 $\mu W$, respectively. But \systemName still saves at least 44\% of the energy compares to light sensor based systems. Furthermore, current light sensor based systems implement an array of light sensors (e.g., 9 in~\cite{kaholokula2016reusing} and 36 in~\cite{venkatnarayan2018gesture}), which definitely incur much higher power consumption.  
 \begin{table}[t]
  \centering
  \small
  \caption{Power consumption comparison.}
  \vspace{-0.15in}
  \label{tab:power analysis}
  \resizebox{3.3in}{!}{
  \begin{tabular}{cccccc}
  \toprule
  \multirow{2}{*}{\textbf{Sensor}} &  \multicolumn{4}{c}{\textbf{Power Consumption (uW)}} &  \multirow{2}{*}{\textbf{Savings}} \\ \cmidrule{2-5}
   & \textbf{MCU}&\textbf{Sensor}&\textbf{Backscatter}   & \textbf{BLE}&    \\ \cmidrule{1-6}
  Solar Cell  & 20.28  & 0 &  0.023 & 31.11   &   \\ \cmidrule{1-6}
  OPT101  &20.28& 39.78&0.023 & 31.11 & 66.2\%/43.6\% \\ 
  \cmidrule{1-6}
  SD3410&20.28&42.18&0.023&31.11  & 67.5\%/45.1\%  \\ 
    \bottomrule 
  \end{tabular}}
  \vspace{-0.1in}
  \end{table}

\section{Related work}
\label{sec:related}

\subsection{Gesture Recognition}
Gesture recognition has been extensively investigated in the literature. Vision based gesture recognition leverages a camera or depth sensor to detect gestures~\cite{izadi2011kinectfusion,howe2000bayesian}. Motion sensor-based approach exploits accelerometers and gyroscopes to track human body/hand movement~\cite{ruiz2011user,xu2012taplogger}. RF signal based gesture recognition systems utilize informations extracted from the RF signal, such as RSS~\cite{abdelnasser2015wigest,8519328}, CSI~\cite{sbirlea2013automatic}, and Doppler shift~\cite{pu2013whole} to recognize different gestures. Acoustic signal based method operates by leveraging Doppler shift of the reflected sound waves caused by gestures~\cite{gupta2012soundwave,pittman2016multiwave}. The underlying principle of light sensor based approach is that different gestures leave distinct shadows that can be captured by an array of photodiodes~\cite{venkatnarayan2018gesture,kaholokula2016reusing,li2017reconstructing,li2015human,li2016practical,li2018self}.

Although such systems achieve excellent gesture recognition accuracy, they actually suffer from some limitations. For example, vision based systems usually incur heavy computation cost and encounter privacy concerns that arise from the sensitive camera data~\cite{sbirlea2013automatic,weinberg2011still}. Moreover, these systems suffer from high energy consumption due to the use of various sensors, like accelerometer, depth sensor, and microphone, which impedes the aim of ubiquitous and perpetual gesture recognition. In contrast, \systemName utilizes the energy harvesting signal for gesture recognition, which not only eliminates sensor-consumed energy but also provides inexhaustible power supply to the IoT device.
% \vspace{-0.2in}
\subsection{Solar Energy Harvesting based Sensing}
Based on the principle that solar energy harvesting signal can be a reflection of the environment light intensity, researchers also utilize the solar cell as a light indicator to perform indoor positioning~\cite{randall2007luxtrace}. In addition, coarse-grained asset localization is achieved by analyzing the harvested energy patterns of a solar panel~\cite{chen2016sunspot}. A recent work~\cite{li2018self} employed an array of photodiodes around a smartwatch for the dual-use of solar energy harvesting as well as gesture recognition. Compared to \cite{li2018self}, a key advantage of \systemName is that it can be seamlessly integrated to the smartwatch screen without impacting its appearance.
 
%Instead of operating in normal photoconductive mode, photodiodes are switched to photovoltaic mode that requires no power supply but generates energy.}

In terms of solar cell based gesture recognition, the most relevant work is~\cite{varshney2017battery}, in which the authors utilized an opaque solar cell to identify three hand gestures. However, \systemName differs in three aspects. First, \cite{varshney2017battery} differentiates three gestures, Swipe, Two Taps, and Four Taps, based on repetitions of a basic gesture, while \systemName recognizes gestures baed on their unique patterns. Second, since transparent solar cell has much lower energy harvesting efficiency compared to the opaque counterpart, its gesture recognition capability was hitherto untested. We demonstrated their gesture recognition potential by prototyping transparent cells and performing practical experiments with them. Third, we developed a theoretical model to investigate the effect of different practical parameters in a solar based gesture recognition system and conducted a comprehensive experiment study to evaluate the gesture recognition performance with different solar cell transparencies and light intensities.

%  	\subfigure[]{
%    		\includegraphics[scale=0.38]{fig/plotPowerConsumption.pdf}	
%    		\label{fig:plotPowerConsumption}}
%    		\vspace{-0.2in}

\section{Conclusion}
\label{sec:conclusion}
We have proposed \systemName, a solar-based gesture recognition system for ubiquitous solar-powered IoTs. Using solar energy harvesting fundamentals and geometric analysis, we derived a model that accurately simulates arbitrary hand gestures and enables estimation of gesture recognition performance under various conditions. Employing real solar cells, both opaque and transparent, we have demonstrated that our system can detect six gestures with 96\% accuracy under typical use scenarios while consuming 44\% less power compared to light sensor based approach. Although we motivated the use case of transparent solar cells on the screens of mobile devices, we have not analyzed the impact of backlight on gesture recognition. Transparent solar cells are still at early stages of research and as such we do not have access to commercially available cells complete with development kits for integrating to IoT development platforms. However, such experiments can be conducted in the near future as soon as transparent cells become commercially available at low cost. When such opportunities arrive, we intend to extend our simulator with capabilities to analyze impact of incident lights from both sides of a transparent solar cell.

%\section*{Acknowledgements}
%nidnd

%For future directions, we will consider: (1) Different from opaque solar cells, transparent solar panel can actually harvest energy from both top layer and bottom layer. In our experiment, hand gestures only modify light absorbed by the top layer. However, considering a transparent solar cell is placed over a smartwatch screen, its bottom layer will absorb energy from screen light as well. Thus, how the changing screen contents (i.e., brightness) affect gesture recognition performance should be further studied. (2) We have investigated the gesture recognition performance with \textit{stable} ambient light at different levels. However, if the light intensity \textit{varies} during a gesture (e.g., under mobility), the corresponding gesture profile should be interfered. Thus, deeper work should be done to investigate such impact and devise the solutions.

\balance
\normalem
\bibliographystyle{unsrt} 
\bibliography{ref_Infocom}

\end{document}